\documentclass[10pt]{article}

\usepackage[numbers]{natbib}

\usepackage[
  a4paper,
  textwidth=126mm,   
  textheight=195mm,  
  centering
]{geometry}

\usepackage{amsmath}
\usepackage{amsfonts}
\usepackage{amssymb}
\usepackage{graphicx}
\usepackage{subcaption}
\usepackage{adjustbox}
\usepackage{wrapfig}
\usepackage{float}

\usepackage{booktabs}
\usepackage{tabularx}
\usepackage{longtable}
\usepackage{multirow}
\usepackage{makecell}
\usepackage{array}

\usepackage[utf8]{inputenc}
\usepackage[T1]{fontenc}
\usepackage{lmodern}
\usepackage{url}
\usepackage{microtype}
\usepackage{xcolor}
\usepackage{xr-hyper}
\usepackage{hyperref}
\hypersetup{hidelinks}
\usepackage[all]{hypcap}
\usepackage{algorithm}
\usepackage{algpseudocode}

\title{Sparse Contextual Coupling Reshapes Diffusion Geometry in Multilayer Hypergraphs}

\author{
Hao Ding \\
Department of Computer Science \\
University of California, Santa Barbara, CA 93106, USA \\
\texttt{hao\_ding@ucsb.edu}
\and
Sanjukta Krishnagopal \\
Department of Computer Science \\
University of California, Santa Barbara, CA 93106, USA \\
\texttt{sanjukta@ucsb.edu}
}

\date{}

\begin{document}

\maketitle

\begin{abstract}
Many complex systems combine dense background structure with sparse contextual
information. In higher-order networks, even a small context-specific layer may
alter global organization because information can propagate through group
interactions rather than only through pairwise edges. We introduce a
diffusion-based framework for analyzing reorganization of diffusion geometry via sparse coupling in multilayer hypergraphs.
Each layer is represented as a weighted hypergraph, layers are coupled through
shared entities, and random walks on the coupled system induce multiscale
diffusion distances between nodes.

We apply this framework to disease-conditioned gene networks by coupling a dense
MSigDB functional gene-set layer to sparse disease-specific DGIdb drug--gene
hypergraphs, with disease-associated drugs selected from DDDB and HumanNet-GSP
used to define external gene weights. Across Bipolar Disorder, Schizophrenia,
Leukemia, and Breast Cancer, the disease-specific layer contains fewer than
\(2\%\) of the genes in the coupled system. Nevertheless, adding this sparse
layer substantially changes diffusion distances and community structure.
Centrality analysis shows that DGIdb-associated genes occupy relatively
influential positions in the MSigDB-derived functional network, providing a
mechanism for their disproportionate effect on diffusion geometry.

The resulting diffusion-derived communities are stable under subsampling and
show coherent post hoc functional enrichment, including signaling and
neurotransmission categories in neuropsychiatric diseases and immune,
translational, and metabolic categories in cancer-associated diseases.
Community-level comparisons further reveal disease similarities not reducible to
direct DGIdb gene overlap. They recover, via data alone, relationships consistent with recent
biomedical evidence, such as similarities between Breast Cancer and Schizophrenia. These results show that sparse contextual layers can induce
interpretable nonlocal changes in higher-order network geometry and provide a
general framework for integrating sparse condition-specific information with
dense functional background structure.
\end{abstract}


\section{Introduction}
\label{sec:introduction}

Many complex systems are organized by interactions among groups of entities rather than by pairwise relationships alone \cite{battiston2020beyond,benson2016higherorder,majhi2022dynamics}. Examples include co-authorship teams, ecological assemblages, protein complexes, biological pathways, drug target sets, and other group-level biological relationships. Pairwise graph representations can be useful approximations, but they necessarily flatten group-level structure and may obscure collective effects that arise only when several entities participate in a common interaction. Hypergraphs provide a natural representation for such higher-order systems by allowing a single edge to connect an arbitrary number of nodes \cite{berge1989hypergraphs}.

Real systems are also rarely homogeneous. Instead, they often contain multiple layers that encode different relation types, levels of confidence, or contexts, forming a `multilayer network' \cite{multilayer,krishnagopal2023topology,krishnagopal2020trajectory}. A biological network, for example, may combine curated pathway memberships, gene--gene functional associations, drug targets, disease associations, and molecular measurements. These layers can differ dramatically in size and density. A dense background layer may encode broadly useful functional structure, whereas a sparse layer may provide the contextual information most relevant to a specific condition. 
We ask the question: when can a small contextual layer have a large nonlocal effect on global diffusion distances and community organization? This question is important in many networked systems. For example, many network-based biomedical analyses \cite{krishnagopal2022stroke,krishnagopal2020parkinson} begin with sparse disease annotations or seed genes and propagate, smooth, or prioritize them over a much larger molecular interaction or functional network, effectively treating the disease-specific information as a sparse contextual signal on a fixed background network \cite{vanunu2010associating,cowen2017network,menche2015uncovering,kim2022network}. However, if the contextual nodes occupy central or bridging positions in the background layer, even a small disease-specific layer may substantially redirect diffusion pathways.
This problem is especially challenging in higher-order settings because information can propagate through group interactions rather than only along pairwise edges: a single hyperedge can couple many nodes simultaneously and thereby alter random-walk, contagion, synchronization, and clustering behavior in ways not captured by ordinary graph projections \cite{carletti2020random,battiston2020beyond,majhi2022dynamics,boccaletti2023structure}. 
Recent work has begun to extend these ideas to multilayer and multiplex higher-order systems, including percolation on multiplex hypergraphs, random-walk representations of hypergraph flows, multilayer hypergraph clustering, Laplacian-based learning on multilayer hypergraphs, and multiplex measures for higher-order networks \cite{sun2021higher,eriksson2021choosing,alaluusua2023multilayer,venturini2023laplacian,lotito2024multiplex}. To our knowledge, relatively little prior work has studied how a sparse condition-specific higher-order layer can globally reshape diffusion geometry when coupled to a much denser hypergraph background layer.


Gene regulation, biological pathways, protein complexes, and drug target relationships are naturally higher-order: they involve sets of genes rather than isolated pairwise interactions. Disease biology is also contextual. A set of drug targets associated with a given disease may be small relative to the full functional gene universe, but may nevertheless occupy important positions in the background network. 
This makes disease-conditioned gene analysis an appropriate test case for studying sparse interlayer influence.
In biology, existing hypergraph approaches have been used for several problems including drug--target prediction \cite{ramadan2004hypergraph,tran2012hypergraph}, driver-gene discovery, and disease-gene prioritization \cite{feng2021hypergraph,wang2022driverrwh,deng2024identifying,jin2023general}. Many of these methods either focus on a single data modality, emphasize local connectivity or centrality, or rely on black-box learning models. 
Here, we study this sparse-contextual regime by constructing disease-conditioned
multilayer hypergraphs that couple a sparse drug--gene layer to a dense
functional gene-set layer. For each disease, disease-associated drugs are
identified from DDDB \cite{hussain_diseasedrug_2016}, and their DGIdb target sets define
drug--gene hyperedges \cite{dgidb_paper}. The shared functional background layer is constructed
from MSigDB gene sets \cite{MSigDB_paper}. The two layers are coupled through genes present in both
layers, while HumanNet-GSP provides independent functional weights on genes \cite{HumanNet_paper}.

To analyze the resulting coupled system, we define a Markov diffusion operator
on layer-specific gene copies and compare genes using multiscale diffusion
distances \cite{coifman2005geometric,coifman2006diffusion}.  These distances compare the full random-walk profiles induced by each starting gene, thereby integrating information across many paths, both within and between layers. This makes them well suited for detecting mesoscopic organization in noisy, heterogeneous, and incomplete biological networks. 
We then aggregate layer-specific diffusion profiles back to unique genes, construct a diffusion-derived nearest-neighbor graph, and identify diffusion-derived gene communities using Leiden clustering \cite{traag2019leiden}. This final graph is an affinity graph in diffusion-profile space, not a clique expansion or pairwise projection of the input hypergraph. Importantly, geometric proximity in the diffusion-derived network should not be
interpreted as a direct physical or mechanistic notion of biological distance.
The underlying databases are incomplete, noisy, and heterogeneous, and many
biological relationships are indirect rather than pairwise. Diffusion distance
instead measures similarity of information flow through the coupled multilayer
system: two genes are considered similar when random walks starting from them
spread through similar functional regions and interact with similar collections
of pathways, gene sets, and disease-associated targets across multiple scales.
Consequently, diffusion geometry can capture mesoscale functional similarity even
when direct adjacency or shortest-path distance is biologically ambiguous.
Functional enrichment is performed only after clustering, so the detected communities are determined by diffusion geometry rather than by
enrichment annotations.

The main contributions of this work are as follows.

\begin{enumerate}
    \item \textbf{Sparse contextual coupling in multilayer hypergraphs.}
    We study the regime in which a small, condition-specific hypergraph layer is coupled to a much denser higher-order background layer. In our application, sparse disease-filtered drug--gene hyperedges are coupled to a shared MSigDB functional gene-set hypergraph.

    \item \textbf{A diffusion-geometric analysis of sparse-layer influence.}
    We use multiscale diffusion distances to measure how sparse interlayer coupling perturbs global geometry, rather than treating disease-specific annotations only as labels or seed sets on a fixed network.

    \item \textbf{Evidence for nonlocal amplification by central shared genes.}
    Across four diseases, DGIdb-associated genes make up fewer than 2\% of the coupled system but substantially alter diffusion distances and community structure. Centrality analysis suggests that this effect is amplified because shared disease-associated genes occupy influential positions in the dense functional background.

    \item \textbf{Disease-conditioned communities with post hoc biological interpretation.}
    Diffusion-space clustering yields communities that combine inherited MSigDB structure with disease-specific drug--gene context. Enrichment analysis reveals plausible neuropsychiatric and cancer-related functional categories, while community-level comparisons reveal disease similarities not explained solely by direct DGIdb gene overlap, including strong Bipolar Disorder--Breast Cancer similarity and a Schizophrenia--Breast Cancer association consistent with recent genetic studies \cite{lu2020shared,tang2023epidemiological,solmi2024antipsychotic,kern2024association}.

\end{enumerate}

\section{Data}
\label{sec:data}

We integrate four public resources: the Disease--Drug Database (DDDB), the Drug--Gene Interaction Database (DGIdb), the Molecular Signatures Database (MSigDB), and HumanNet-GSP. Table~\ref{tab:datasets} summarizes their roles in the analysis. 

\begin{table}[H]
\caption{Summary of datasets used for hypergraph construction and interpretation.}
\label{tab:datasets}
\renewcommand{\arraystretch}{1.2}
\small
\centering
\begin{tabularx}{\textwidth}{p{1.5cm}p{1.5cm}p{4.5cm}X}
\toprule
\textbf{Dataset} & \textbf{Type} & \textbf{Description} & 
\textbf{Role in analysis} \\
\midrule
DDDB & Hypergraph & Disease--drug mapping between 324 diseases and 1,321 drugs & Selects drugs associated with each disease \\

DGIdb & Hypergraph & Drug--gene interactions between 4,774 genes and 18,545 drugs & Constructs the disease-filtered drug--gene hypergraph layer \\

MSigDB C2 & Hypergraph & Collection of 7,411 curated gene sets among 21,981 genes & Constructs the dense functional hypergraph layer \\

HumanNet-GSP & Graph & Functional gene network encoding 260,962 gene--gene associations of 8,779 genes & Provides external weights for genes in both hypergraph layers \\
\bottomrule
\end{tabularx}
\end{table}

For each disease, DDDB is used to identify associated drugs \cite{hussain_diseasedrug_2016}. DGIdb aggregates curated drug--gene interactions from publications and expert resources \cite{dgidb_paper}; the DGIdb target sets of the DDDB-associated drugs then define the disease-specific drug--gene hyperedges. MSigDB provides gene sets representing pathways and other functional programs. For our analysis, we used the MSigDB C2 collection, a curated collection of canonical pathway gene sets and chemical/genetic perturbation signatures \cite{MSigDB_paper}. These gene sets define the dense background functional layer, which is common across diseases. Both DGIdb and MSigDB are aggregative knowledge bases that synthesize evidence from numerous studies and source resources. This broad coverage reduces dependence on any single experiment, although biases shared across the biomedical literature may remain. HumanNet-GSP provides a weighted functional gene network integrating evidence from multiple biological sources \cite{HumanNet_paper}. We use the HumanNet-GSP degree of each gene as an external weight in both hypergraph layers, thereby emphasizing genes with broader independent functional support.

\section{Methods}
\label{sec:methods}

\subsection{Overview}
\begin{figure}[H]
    \centering
    \includegraphics[width=\textwidth,height=0.34\textheight]{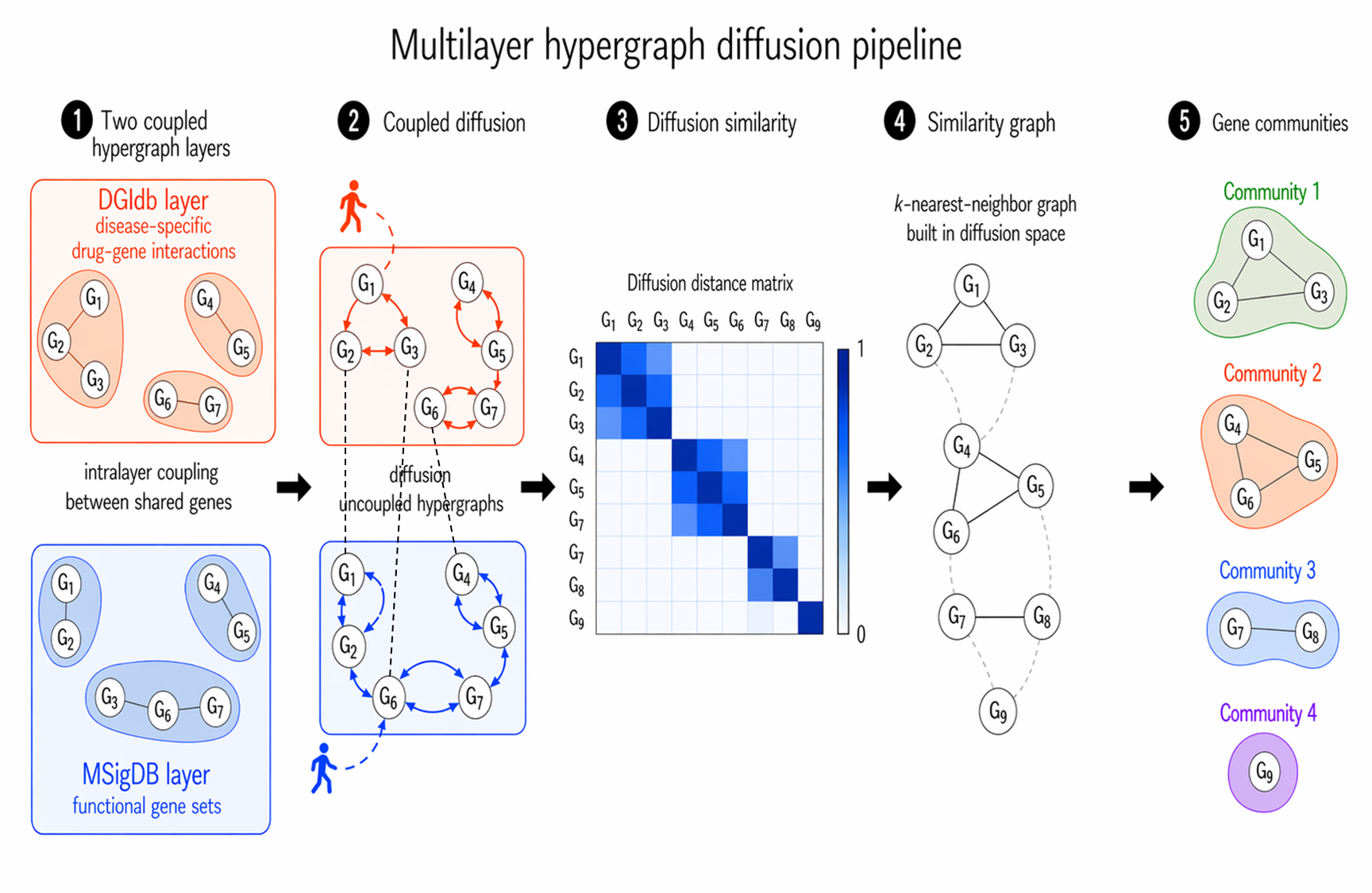}
    \caption{Overview of the analysis pipeline. Disease-associated drugs define a sparse DGIdb drug--gene hypergraph layer, while MSigDB defines a dense functional gene-set hypergraph layer. The two layers are coupled through shared genes. A multilayer Markov diffusion process induces multiscale diffusion distances, from which we construct a $k$NN graph, identify communities, and perform downstream functional interpretation.
    \newline
    \textbf{Alt text}: Left-to-right workflow showing coupled hypergraph layers for the DGIdb and the MSigDB. Diffusion within and between layers produces a gene distance heatmap, which defines a nearest-neighbor similarity graph. The graph is partitioned into four gene communities.}
    \label{fig:pipeline}
\end{figure}

For each disease, we construct a two-layer hypergraph consisting of a sparse
DGIdb drug--gene layer and a dense MSigDB functional gene-set layer shared
across diseases. The MSigDB layer provides broad functional background structure,
whereas the DGIdb layer provides disease-specific pharmacological context. Genes
appearing in both layers serve as coupling points: a random walk can move through
hyperedges within either layer and can pass between layers at shared genes. Thus,
sparse disease-specific information can propagate into the dense functional
background.

We compare genes using the resulting multistep diffusion profiles. Two genes are
diffusion-similar if random walks starting from them spread through the coupled
system in similar ways and reach similar regions over time. After aggregating
layer-specific copies back to unique genes, we compute multiscale diffusion
distances, construct a diffusion-derived nearest-neighbor graph where genes with similar diffusion profiles are strongly connected to each other, and identify
communities. We then compare the coupled geometry with the MSigDB-only baseline,
quantify how interlayer coupling changes diffusion distances, and perform post
hoc enrichment analysis. The full pipeline is summarized in
Figure~\ref{fig:pipeline}.

\subsection{Multilayer hypergraph construction}
\label{sec:hypergraph_construction}
Let $b$ denote a disease. For each disease \(b\), we construct a two-layer hypergraph
\begin{equation}
\mathcal{H}_{b}=\{\mathcal{H}_{b}^{(D)},\mathcal{H}^{(M)}\},
\end{equation}
where $\mathcal{H}_{b}^{(D)}$ is the disease-specific DGIdb layer (denoted by $D$ and disease $b$) and $\mathcal{H}^{(M)}$ is the MSigDB functional layer (denoted by $M$). The DGIdb layer varies with $b$, whereas the MSigDB layer is common across diseases.

Each layer is represented by a weighted incidence matrix. Let \(\{e_j\}\) denote the hyperedges in layer \(L\), where each hyperedge
corresponds either to a DGIdb drug-target set or an MSigDB gene set.
The weighted incidence matrix \(H^{(L)}\) is defined by
\begin{equation}
H^{(L)}_{ij} =
\begin{cases}
 w(v_i), & v_i \in e_j,\\
 0, & \text{otherwise.}
\end{cases}
\end{equation}
Here $w(v_i)$ is a node weight derived from HumanNet-GSP. Specifically, let \(\mathcal N=(V_{\mathcal N},E_{\mathcal N})\) denote the HumanNet-GSP functional gene network, the weight for a node \(v_i\) is
\begin{equation}
w(v_i)=
\begin{cases}
\deg_{\mathrm{\mathcal N}}(v_i), & v_i \in V_{\mathrm{\mathcal N}},\\
0.01, & v_i \notin V_{\mathrm{\mathcal N}}.
\end{cases}
\end{equation}
This weighting scheme emphasizes genes in HumanNet, i.e. genes with independent functional evidence, while retaining genes absent from HumanNet with small weight. Table~\ref{tab:human_net_percent} reports the percentage of genes in the set of DGIdb and MSigDB genes that also appear in the HumanNet-GSP gene set $V_{\mathcal{N}}$ for each disease. 

\begin{table}[ht]
\caption{Percentage of genes in the multilayer hypergraph that intersect genes in HumanNet-GSP.}
\label{tab:human_net_percent}
\centering
\setlength{\tabcolsep}{6pt}
\renewcommand{\arraystretch}{1.2}
\begin{tabular}{lccccc}
\toprule
 & Bipolar & Schizo. & Leukemia & Breast Ca. & MSigDB \\
\midrule
HumanNet overlap (\%) & 68.52 & 71.08 & 80.30 & 72.94 & 39.78 \\
\bottomrule
\end{tabular}
\end{table}

See Appendix A1 for more details on the construction of the multilayer hypergraph.

\subsection{Layer-wise hypergraph diffusion}
\label{sec:single_layer_diffusion}

For layer $L$, define the node and hyperedge degree vectors
\begin{equation}
d_v^{(L)}(i)=\sum_j H^{(L)}_{ij},
\qquad
 d_e^{(L)}(j)=\sum_i H^{(L)}_{ij},
\end{equation}
and let $D_v^{(L)}$ and $D_e^{(L)}$ be the corresponding diagonal matrices. We define a row-stochastic transition matrix
\begin{equation}
A^{(L)} = {D_v^{(L)}}^{-1} H^{(L)} {D_e^{(L)}}^{-1} {H^{(L)}}^\top .    
\end{equation}

This operator is the standard incidence-based hypergraph random walk, avoiding an explicit clique expansion. At each step, a walker selects an incident hyperedge according to the weighted incidence structure and then selects a gene within that hyperedge using hyperedge-normalized weights. In the unweighted form, the contribution of hyperedge $e$ to a transition from node $i$ to a particular member contains the factor $1/(d_v^{(L)}(i)|e|)$. Summing over its $|e|$ possible destinations cancels this cardinality factor, so increasing hyperedge size does not automatically multiply the total transition probability assigned to that hyperedge. The weighted hyperedge-degree normalization in our implementation serves the same purpose. Since the current gene belongs to the selected hyperedge, the walk naturally includes a positive self-transition probability.
\subsection{Multilayer coupling and Markov operator}
\label{sec:layer_couping}
The two layers are coupled only through genes appearing in both DGIdb and MSigDB. We treat a gene appearing in both layers as a \emph{supra-node} having two layer-specific copies. Let $\beta\in[0,1]$ denote the interlayer transition probability. For a DGIdb-layer gene $i$ with corresponding MSigDB-layer copy $i'$, the interlayer transition from DGIdb to MSigDB consists of jumping to the MSigDB copy and then taking one diffusion step in the MSigDB layer. Thus,
\begin{equation}
{(C_{12})}_{i:} =
\begin{cases}
A^{(M)}_{i':}, & \text{if gene } i \text{ in DGIdb has a copy } i' \text{ in MSigDB},\\
\mathbf{0}, & \text{otherwise.}
\end{cases}
\end{equation}
Similarly,
\begin{equation}
{(C_{21})}_{i:} =
\begin{cases}
A^{(D)}_{i':}, & \text{if gene } i \text{ in MSigDB has a copy } i' \text{ in DGIdb},\\
\mathbf{0}, & \text{otherwise.}
\end{cases}
\end{equation}
Let \(G_D\) and \(G_M\) denote the gene sets in the DGIdb and MSigDB layers,
respectively. Let $B^{(D)} \in \mathbb R^{|G_D| \times |G_D|}$ and $B^{(M)} \in \mathbb R^{|G_M| \times |G_M|}$ be diagonal matrices whose entries are $\beta$ for genes with cross-layer copies and $0$ otherwise. The coupled Markov operator is
\begin{equation}
P =
\begin{bmatrix}
(I-B^{(D)})A^{(D)} & B^{(D)}C_{12}\\
B^{(M)}C_{21} & (I-B^{(M)})A^{(M)}
\end{bmatrix}.    
\end{equation}
Thus, at a shared gene, the walker either remains in the current layer and diffuses there with probability $1-\beta$, or switches to the other layer and then diffuses in that layer with probability $\beta$. Genes without cross-layer copies have no interlayer transition and therefore remain within their original layer. Since each row is a convex combination of row-stochastic transition probabilities, $P$ is row-stochastic.

\subsection{Diffusion distances and the diffusion-derived nearest-neighbor graph}
\label{sec:diffusion_distance_and_knn_graph}

We use diffusion distance to measure gene similarity because it compares the full multistep diffusion profiles induced by the coupled Markov operator, thereby capturing both local hyperedge membership and broader connectivity through the multilayer system.

For a diffusion time \(t\), the \(i\)th row of \(P^t\) gives the probability
distribution of a random walker after \(t\) steps, conditioned on starting from
supra-node \(i\). We call this the $t$-step diffusion profile of $i$. Two genes are considered similar if their diffusion profiles are similar, i.e. if random walks started from the two genes spread through the coupled system in similar ways.
Because a gene appearing in both DGIdb and MSigDB has two layer-specific copies, we first aggregate diffusion profiles from supra-nodes back to unique genes. Let $A_r$ and $A_c$ denote the row and column aggregation matrices defined in Appendix A2. The aggregated $t$-step diffusion profile matrix is
\begin{equation}
\widetilde P^t = A_r P^t A_c.    
\end{equation}

Here, right multiplication by $A_c$ combines probability mass assigned to layer-specific copies of the same destination gene, while left multiplication by $A_r$ combines diffusion profiles associated with duplicated source-gene copies. The squared diffusion distance between genes $i$ and $j$ is then
\begin{equation}
D_t^2(i,j)=\| (\widetilde P^t)_i-(\widetilde P^t)_j\|_2^2.
\end{equation}

Thus, $D_t^2(i,j)$ is small when walks starting from $i$ and $j$ reach similar regions of the multilayer hypergraph after $t$ steps. Because different diffusion times capture different spatial scales, we average squared distances over multiple diffusion times in the set $T$ to capture multiple scales:
\begin{equation}
\overline D^2(i,j)=\frac{1}{|T|}\sum_{t\in T}D_t^2(i,j).
\end{equation}
Shorter times emphasize local neighborhoods, while longer times reflect broader connectivity.

Finally, we convert distances to similarities using a Gaussian kernel, which turns proximity in diffusion space into weights of edges in a gene-gene network. We retain the $k$ nearest neighbors of each gene in diffusion space, symmetrize the resulting graph, and use the resulting $k$-nearest-neighbor graph for community detection. An edge in this graph does not indicate that two genes co-occur in a hyperedge, nor is its weight determined by their number of shared hyperedges. It indicates that the genes occupy nearby positions in the multiscale geometry induced by diffusion through the complete multilayer hypergraph. Applying standard downstream algorithms to similarities generated from richer underlying structures is common in diffusion maps, kernel methods, and manifold learning. See Appendix A4 for more details on the construction of the $k$-nearest-neighbor graph.

\subsection{Community detection and enrichment analysis}
\label{sec:community_detection_and_enrichment_analysis}

Diffusion-derived communities are detected by applying the Leiden algorithm \cite{traag2019leiden}
to the diffusion-derived \(k\)NN graph. Leiden optimizes modularity on the
weighted graph and does not require the number of communities to be specified in
advance. Leiden therefore clusters the geometry generated by the higher-order random walk; it does not directly partition the input hypergraph. We retain communities above a minimum size threshold and, for enrichment
analysis, genes whose within-community
weighted degree passes a z-score threshold.

Functional interpretation is performed post hoc. For each detected community,
we test whether its genes are enriched for known biological pathways, Gene
Ontology terms, or functional categories relative to a suitable background gene
set. Enrichr \cite{kuleshov2016enrichr} computes a Fisher exact test for overlap between the community gene set
and each annotated term, followed by Benjamini--Hochberg correction for multiple
testing. We retain only terms whose adjusted \(p\)-value is below a specified
threshold and whose overlap with the community exceeds a minimum fraction of the
term gene set.  Thus, enrichment is used only to interpret diffusion-derived
communities, not to define them. Details are provided in
Appendix A5.

\section{Results}
\label{sec:results}
\textbf{Parameters: }unless otherwise stated we use interlayer transition probability $\beta=0.35$, diffusion times $T = \{2,4,6,8\}$, $k=400$ nearest neighbors, Leiden resolution $\gamma=1.3$, minimum community size $s=100$, centrality threshold $z=0$, adjusted $p$-value threshold $\alpha=10^{-5}$, and overlap threshold $\tau=0.1$.

\subsection{Disease-specific layers are sparse but structurally central}
\label{sec:centrality_analysis}

For each disease, we construct a two-layer hypergraph from DGIdb, MSigDB, and HumanNet-GSP. Table~\ref{tab:networkstats} summarizes network statistics for four representative diseases. The MSigDB layer contains roughly 22,000 genes and 7,411 hyperedges, whereas the disease-specific DGIdb layers (with disease-associated drugs selected from DDDB and target genes obtained from DGIdb) contain only 198--377 genes depending on the disease and 5--16 hyperedges. Thus, \textbf{fewer than 2\% of genes are shared} between the disease-specific and functional layers.

\begin{table}[h]
\caption{Summary of multilayer hypergraph statistics. The DGIdb layer is disease-specific, while the MSigDB layer is common across diseases. The shared genes column reports the size of the DGIdb gene set relative to the unique gene set in the coupled system. }
\label{tab:networkstats}
\centering
\setlength{\tabcolsep}{2pt} 
\begin{tabular}{lcccc}
\toprule
 & Bipolar Disorder & Schizophrenia & Leukemia & Breast Cancer \\
\midrule
Genes ($|V|$) & 21991 & 21988 & 21984 & 21987\\
DGIdb genes ($|V_b^{(D)}|$) & 359 & 249 & 198 & 377\\
MSigDB genes ($|V^{(M)}|$) & 21981 & 21981 & 21981 & 21981\\
Shared genes (\%) & 1.59 & 1.10 & 0.89 & 1.69\\
DGIdb hyperedges ($|E_b^{(D)}|$) & 16 & 11 & 5 & 10\\
MSigDB hyperedges ($|E^{(M)}|$) & 7411 & 7411 & 7411 & 7411\\
\bottomrule
\end{tabular}
\end{table}

\begin{figure}[H]
    \centering
    \includegraphics[width=\linewidth]{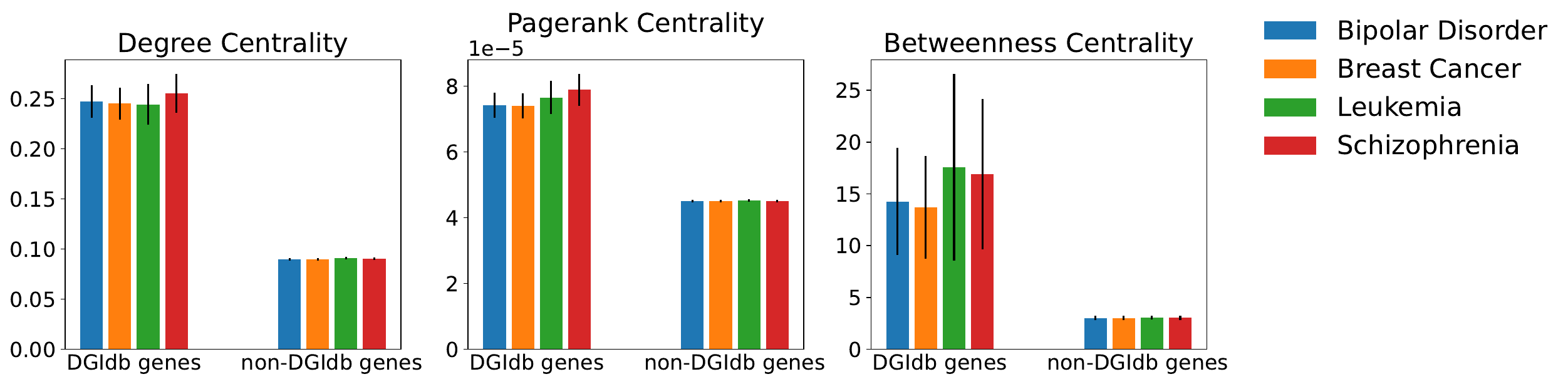}
    \caption{Comparison of mean degree, PageRank, and betweenness centrality. Genes are separated into DGIdb-associated genes that are represented in the DGIdb layer for that disease, and non-DGIdb genes. Results are presented for Bipolar Disorder, Breast Cancer, Leukemia, and Schizophrenia. Centralities are computed on the MSigDB-derived functional network, and DGIdb-associated genes show higher mean centrality across all three measures. Error bars denote standard deviation. AUC and Mann--Whitney statistics are reported in Table~\ref{tab:centrality_auc}.
    \newline
    \textbf{Alt text}: Three grouped bar charts with error bars compare degree, PageRank, and betweenness centrality between DGIdb genes and other genes on the MSigDB layer across bipolar disorder, breast cancer, leukemia, and schizophrenia. DGIdb genes consistently have higher centrality.
    }
    \label{fig:dgidb_centrality_msigdb_adjacency}
\end{figure}

Despite this sparsity, DGIdb genes occupy nonrandom positions in the background
functional network. Figure~\ref{fig:dgidb_centrality_msigdb_adjacency} compares
degree, PageRank, and betweenness centrality in the MSigDB-derived functional
graph for genes that are represented in the DGIdb layer versus genes that are
not. These centrality measures capture complementary notions of influence: degree
centrality measures local connectivity, PageRank emphasizes genes connected to
other well-connected genes, and betweenness centrality identifies genes that lie
on many shortest paths and may therefore act as bridges between functional
regions of the network.
Across diseases, DGIdb-associated genes have higher centrality by all three
measures, suggesting that the sparse disease-specific layer is coupled to the
functional background through relatively influential genes. For computational
tractability, betweenness centrality was computed after transforming similarities
to distances and retaining up to \(k\) shortest paths.

To assess this difference statistically, we performed one-sided Mann--Whitney
\(U\) tests comparing centrality values for DGIdb-associated genes against
non-DGIdb genes. As shown in Table~\ref{tab:centrality_auc}, DGIdb-associated
genes have significantly higher centrality for every disease and every
centrality measure. The corresponding AUC values are consistently above \(0.5\),
indicating that a randomly chosen DGIdb-associated gene is more likely to have
higher centrality than a randomly chosen non-DGIdb gene. However, the AUC values
indicate moderate separation rather than a uniform shift of the entire
distribution. Together with the larger mean centralities in
Figure~\ref{fig:dgidb_centrality_msigdb_adjacency}, this suggests that the
effect is amplified by a subset of DGIdb-associated genes with especially large
centrality values.

\begin{table}[ht]
\centering
\caption{One-sided Mann--Whitney test comparing DGIdb and non-DGIdb gene centralities for Bipolar Disorder, Schizophrenia, Leukemia, and Breast Cancer. The reported $p$-value tests whether DGIdb genes tend to have larger centrality values than non-DGIdb genes. The AUC is $U/(n_{\mathrm{DGIdb}}n_{\mathrm{non\text{-}DGIdb}})$, where $U$ is the Mann--Whitney statistic, and represents the probability that a randomly selected DGIdb gene has higher centrality than a randomly selected non-DGIdb gene.}
\label{tab:centrality_auc}
\begin{tabular*}{0.85\textwidth}{@{\extracolsep{\fill}} llcc}
\toprule
Disease & Centrality & AUC & $p$-value \\
\midrule
Bipolar Disorder & Degree      & 0.668 & $2.01 \times 10^{-30}$ \\
                 & PageRank    & 0.633 & $9.38 \times 10^{-20}$ \\
                 & Betweenness & 0.583 & $3.91 \times 10^{-33}$ \\
\midrule
Schizophrenia    & Degree      & 0.676 & $6.22 \times 10^{-24}$ \\
                 & PageRank    & 0.645 & $8.98 \times 10^{-17}$ \\
                 & Betweenness & 0.603 & $3.96 \times 10^{-35}$ \\
\midrule
Leukemia         & Degree      & 0.677 & $1.29 \times 10^{-19}$ \\
                 & PageRank    & 0.647 & $3.77 \times 10^{-14}$ \\
                 & Betweenness & 0.607 & $1.20 \times 10^{-30}$ \\

\midrule
Breast Cancer    & Degree      & 0.667 & $9.09 \times 10^{-32}$ \\
                 & PageRank    & 0.633 & $1.22 \times 10^{-20}$ \\
                 & Betweenness & 0.584 & $1.88 \times 10^{-35}$ \\
\bottomrule
\end{tabular*}
\end{table}

Together, these results provide a structural mechanism for sparse-layer influence. Although disease-associated genes are few in number, across all four diseases studied, they tend to occupy central or bridging positions in the dense functional network. As a result, interlayer diffusion through these genes can affect regions of the functional layer far beyond the immediate set of shared nodes.

\subsection{Sparse interlayer coupling reshapes diffusion geometry}
\label{sec:sparse_coupling}

We evaluate how adding sparse disease-specific coupling dramatically changes the diffusion geometry. 
To quantify how the strength of interlayer coupling controls the diffusion
geometry, we vary \(\beta\) and compute \(R_\beta\), the log-ratio between each
gene's average diffusion distance to DGIdb-associated genes at coupling strength \(\beta\)
and the corresponding distance in the uncoupled baseline. Formally, we define
\begin{equation}
R_\beta(i)=\log_{10}\left(\frac{\overline d_\beta(i)+\varepsilon_\beta}{\overline d_0(i)+\varepsilon_\beta}\right),
\end{equation}
where $\overline d_\beta(i)$ is the average squared diffusion distance from gene $i$ to all DGIdb genes at coupling strength $\beta$, and
\begin{equation}
\varepsilon_\beta=10^{-8}\operatorname{median}(\overline d_0\cup \overline d_\beta)
\end{equation}
is a negligible scale-adaptive offset.

Intuitively,
\(R_\beta\) measures how much closer or farther a gene becomes, in diffusion
space, to the disease-associated region of the network once cross-layer
transitions are introduced. Negative values indicate that diffusion paths bring
the gene closer to DGIdb-associated genes than in the MSigDB-only baseline,
while positive values indicate increased separation. 

\begin{figure}[H]
    \centering
    \includegraphics[width=\linewidth]{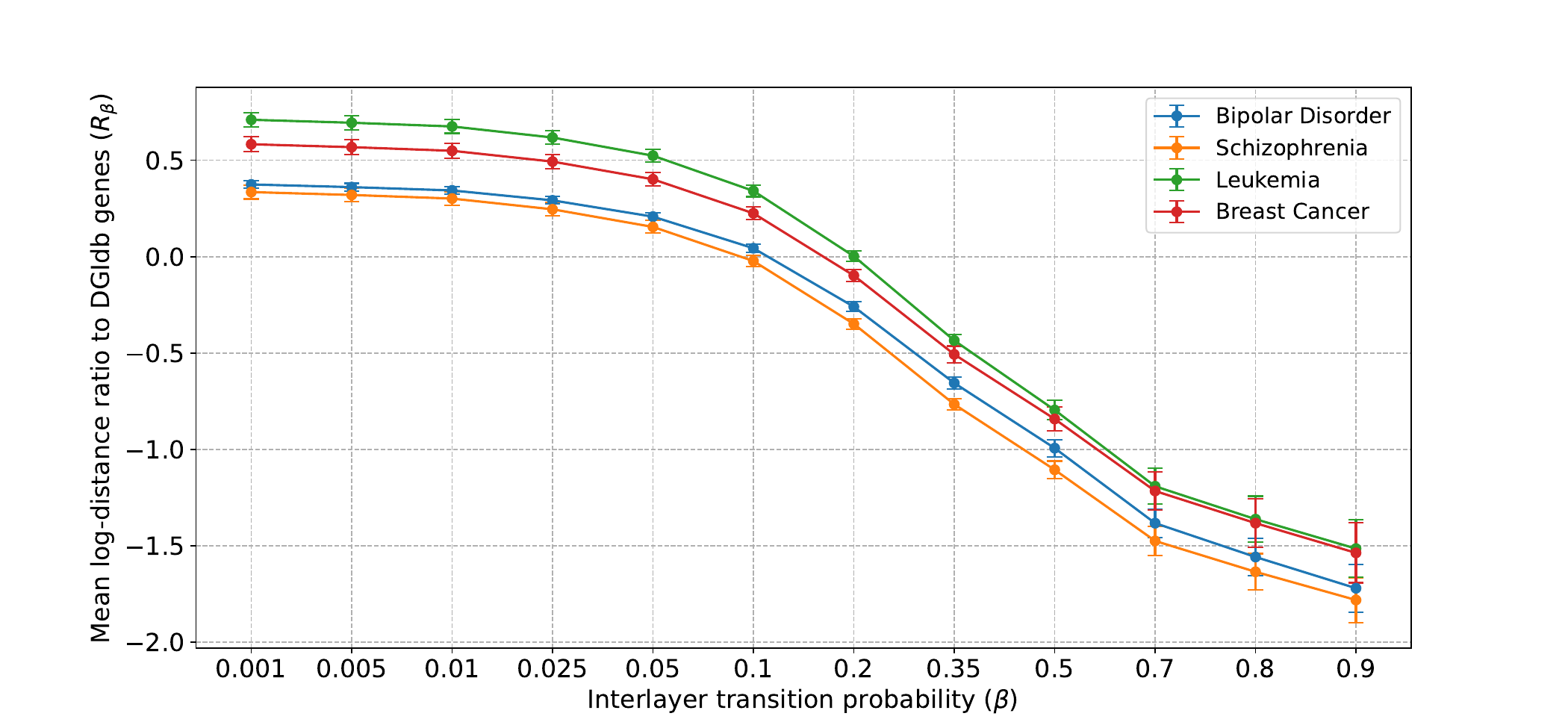}
    \caption{Mean and standard deviation of $R_\beta$ over 2028 sampled genes for Bipolar Disorder, Schizophrenia, Leukemia, and Breast Cancer. The sample was constructed by selecting 30 DGIdb genes and 2000 MSigDB genes without replacement, yielding 2028 distinct genes. Across diseases, the mean decreases smoothly below zero as $\beta$ increases and begins to flatten at larger values, indicating that stronger coupling brings sampled genes closer to DGIdb-associated genes with diminishing additional change at high coupling. The horizontal axis is evenly spaced for visualization.
    \newline
    \textbf{Alt text}: Line plot with error bars showing mean $R_\beta$ values versus interlayer transition probability for bipolar disorder, schizophrenia, leukemia, and breast cancer. All curves decrease as transition probability increases, crossing zero around one tenth to one fifth.}
    \label{fig:mean_and_stdev_R_bipolar}

\end{figure}

As shown in
Figure~\ref{fig:mean_and_stdev_R_bipolar}, across all four diseases, average \(R_\beta\) decreases monotonically
as \(\beta\) increases. Thus, stronger coupling makes genes in the functional
MSigDB layer, on average, closer to disease-associated DGIdb genes. This effect
is visible even for small positive values of \(\beta\), indicating that the
sparse DGIdb layer acts as a nonlocal perturbation of the diffusion geometry
rather than merely adding a small number of local edges. The slight flattening of the curve at larger \(\beta\) suggests that once cross-layer transitions occur frequently
enough, additional coupling produces diminishing geometric changes.

We use $\beta=0.35$ as a prespecified representative value from the stable intermediate regime in Figure~\ref{fig:mean_and_stdev_R_bipolar}, rather than as a data-fitted optimum. Row normalization of the transition matrix already accounts for differences in local weighted density, so choosing $\beta$ from the raw ratio of global layer densities would impose a second density compensation without a direct probabilistic interpretation. The response changes smoothly across neighboring intermediate values, indicating that the substantive conclusions are not tied to a finely tuned choice. More generally, $\beta$ should be selected through a prespecified sensitivity analysis, and conclusions should be reported when they remain stable across a reasonable neighborhood of the chosen value.

Together, these results suggest that sparse disease-specific coupling reshapes
the broader diffusion geometry of the multilayer system rather than producing
only localized perturbations near shared genes. The full distributions of
diffusion distances to DGIdb-associated genes, across different values of
\(\beta\) and across diseases, are shown in
Appendix A9.

\subsection{Degree-preserving configuration-model baseline}
\label{sec:configuration_model_baseline}

To test whether the observed geometric change could be explained simply by adding a sparse layer with the same degree structure, we replaced the Bipolar Disorder DGIdb drug--gene graph with 20 degree-preserving configuration-model randomizations while keeping the MSigDB background and all other model choices fixed. For each Bipolar Disorder DGIdb gene $i$ and diffusion time $t\in T$, we compared its complete diffusion profile after coupling to layer $G$, $x_{G,i}^{(t)}$, with its profile in the fixed higher-order background before DGIdb coupling, $x_{\mathrm{ref},i}^{(t)}$, using
\begin{equation}
\delta_{G,i}^{(t)}=
\frac{\left\lVert x_{G,i}^{(t)}-x_{\mathrm{ref},i}^{(t)}\right\rVert_2}
{\left\lVert x_{\mathrm{ref},i}^{(t)}\right\rVert_2+10^{-12}}.
\end{equation}
We averaged this normalized displacement over diffusion times for each gene and took the median across the Bipolar Disorder DGIdb genes. The observed layer gave $S_{\mathrm{real}}=0.8853$. The 20 null values were
\begin{multline*}
0.5058, 0.5072, 0.5105, 0.5019, 0.4997, 0.5084, 0.5031, 0.5046, 0.5049, 0.5028,\\
0.5038, 0.5050, 0.5035, 0.5054, 0.5029, 0.5093, 0.5085, 0.5042, 0.5038, 0.5047.
\end{multline*}
Their mean was $0.5050$ (range $0.4997$--$0.5105$), making the observed displacement $75.3\%$ larger than the null mean. None of the randomized layers reached the observed value; the one-sided finite-sample Monte Carlo result was $p=0.0476$. Thus, configuration models retaining the observed degree structure do not reproduce the magnitude of diffusion reorganization found in the real network.

\subsection{Communities emerging from diffusion geometry}
\label{sec:communities}
We next ask whether the changes in diffusion geometry induced by sparse
disease-specific coupling are reflected in mesoscale community structure.
Community detection is applied to the diffusion-derived \(k\)NN graph, so genes
are grouped according to similarity of their multiscale random-walk profiles
rather than direct hyperedge overlap alone. Thus, a community represents a set
of genes that occupy similar positions in the coupled diffusion geometry: random
walks initialized from these genes tend to spread through similar functional
regions and disease-associated contexts.

\begin{figure}[p]
\centering
\vspace*{-8em}

\begin{subfigure}[t]{0.39\linewidth}
  \centering
  \includegraphics[width=\linewidth]{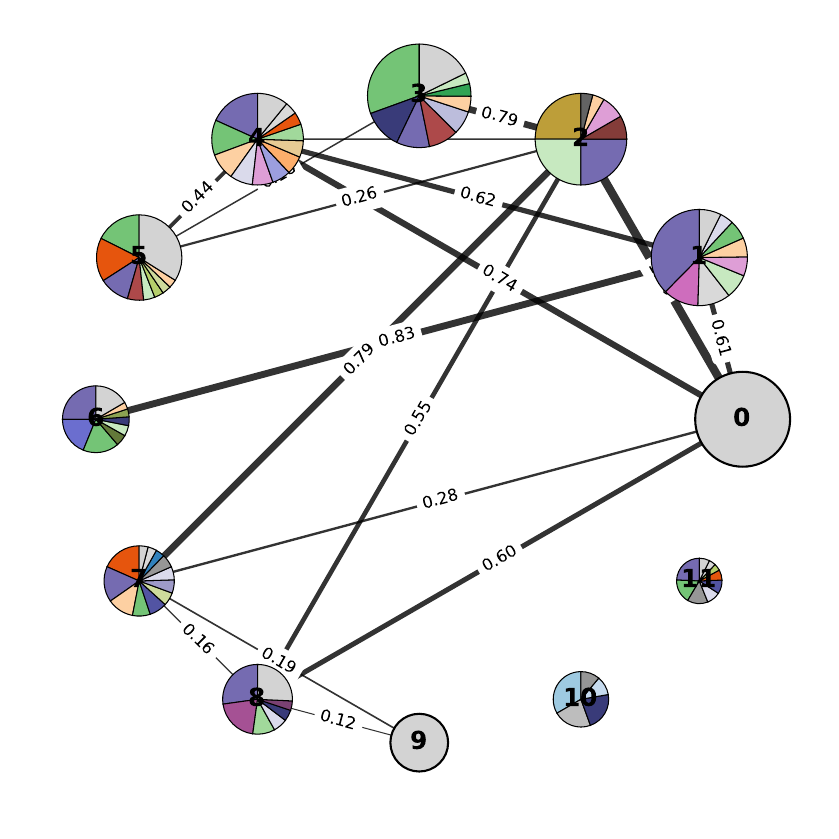}
  \caption{Bipolar Disorder}
\end{subfigure}\hfill
\begin{subfigure}[t]{0.39\linewidth}
  \centering
  \includegraphics[width=\linewidth]{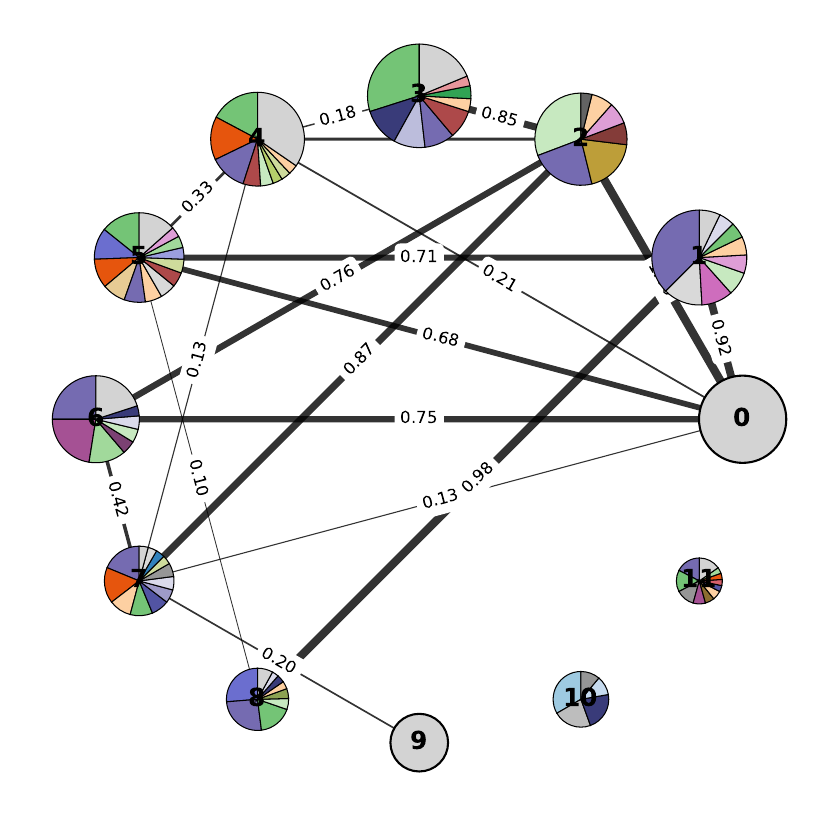}
  \caption{Schizophrenia}
\end{subfigure}

\vspace{0.2em}

\begin{subfigure}[t]{0.39\linewidth}
  \centering
  \includegraphics[width=\linewidth]{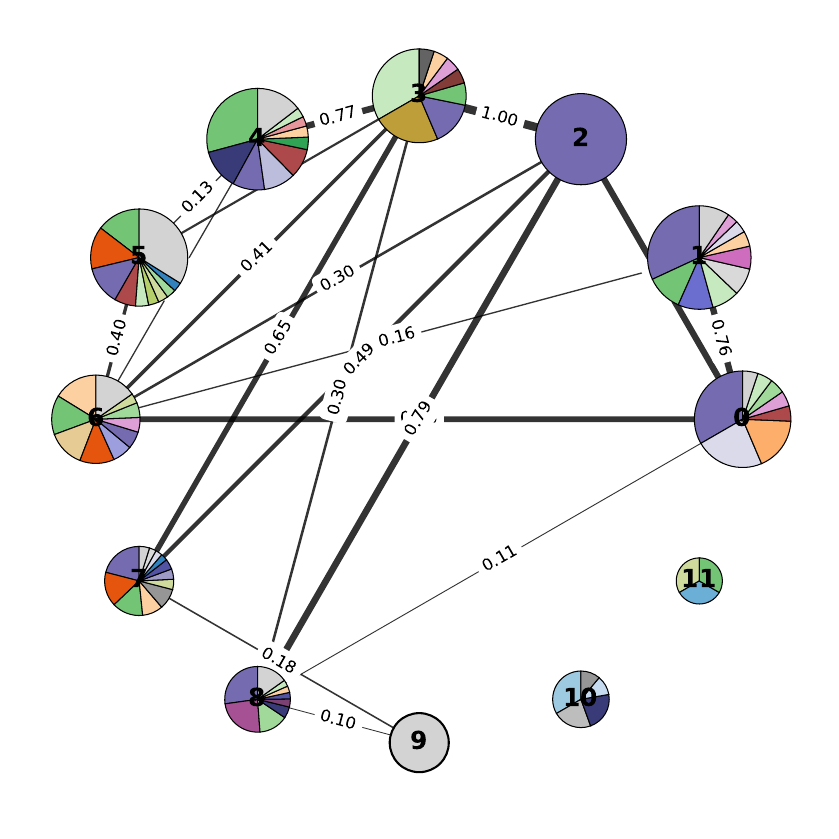}
  \caption{Leukemia}
\end{subfigure}\hfill
\begin{subfigure}[t]{0.39\linewidth}
  \centering
  \includegraphics[width=\linewidth]{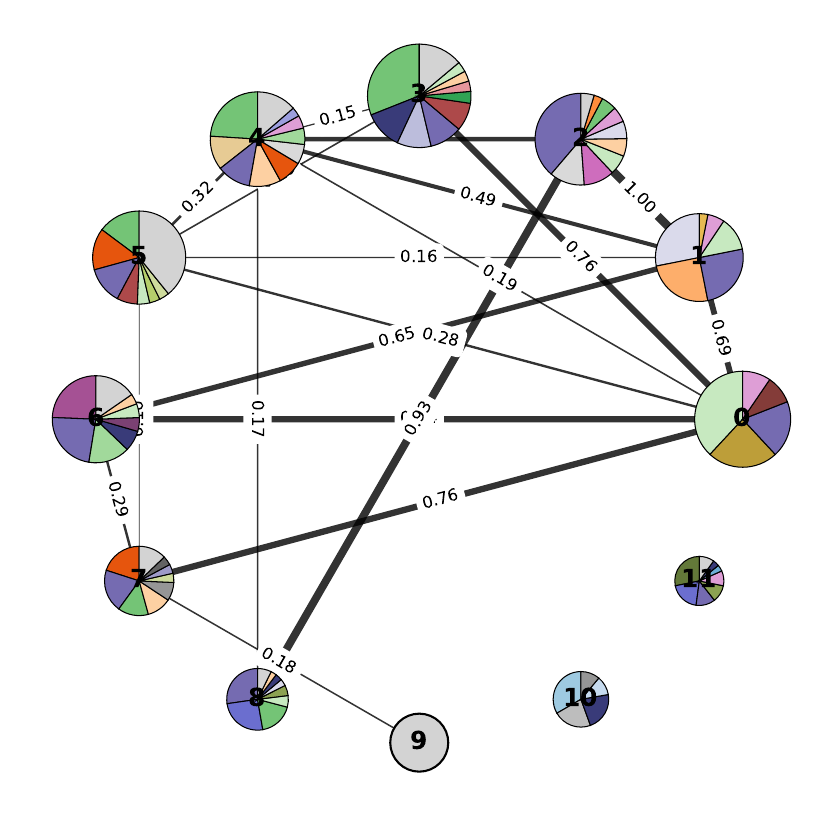}
  \caption{Breast Cancer}
\end{subfigure}

\vspace{0.2em}

\begin{subfigure}[t]{0.39\linewidth}
  \centering
  \includegraphics[width=\linewidth]{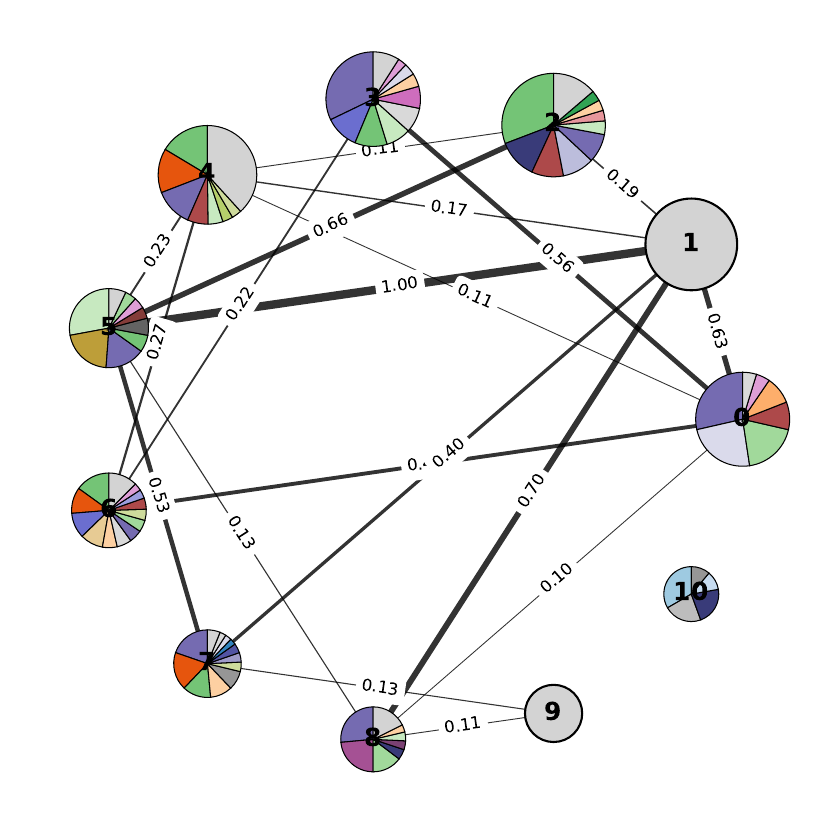}
  \caption{MSigDB-only Baseline}
\end{subfigure}\hfill
\begin{subfigure}[t]{0.55\linewidth}
  \centering
  \includegraphics[width=\linewidth]{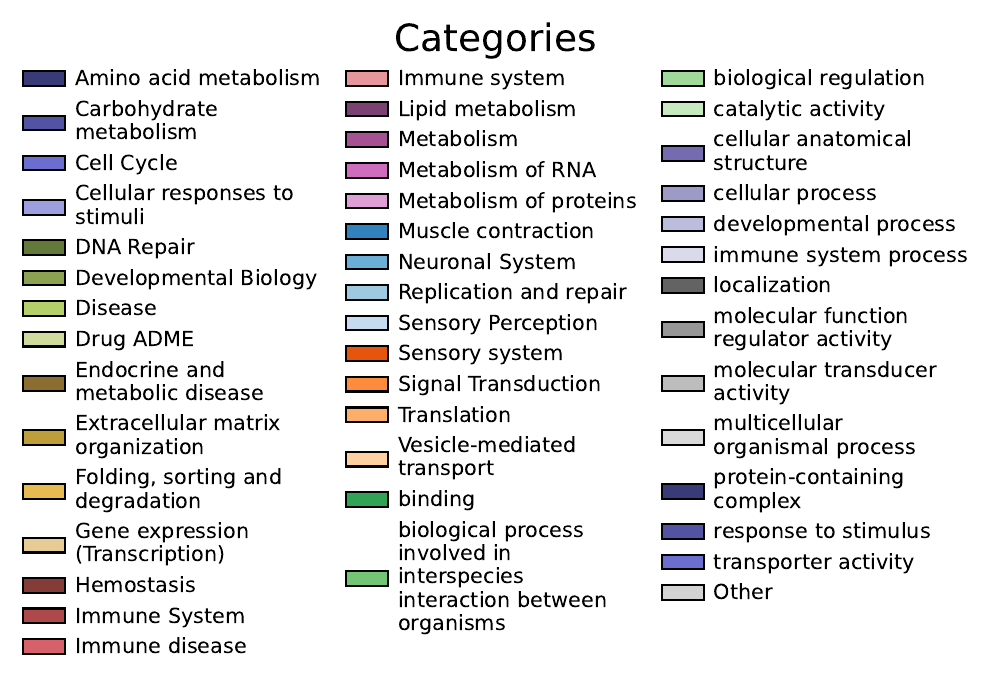}
\end{subfigure}

\caption{Diffusion-derived gene communities for (a) Bipolar Disorder, (b) Schizophrenia, (c) Leukemia, (d) Breast Cancer, and (e) the MSigDB-only baseline. Each node is a Leiden community in the diffusion-derived $k$NN graph; node size indicates the number of genes, pie-chart colors show enriched functional categories, and edge width indicates mean inter-community similarity (Appendix A6). Categories representing fewer than \(3\%\) of retained terms are grouped as ``Other''. The disease-conditioned panels retain broad background organization while differing in composition and connectivity.
\newline
\textbf{Alt text}: Five community networks compare four disease models with a MSigDB-only baseline. Large, interconnected communities coexist with smaller peripheral or isolated communities. Pie-chart proportions representing enriched functional categories and edge widths representing inter-community similarity vary across panels, while the broad network organization remains similar.
}
    
\label{fig:community_detection_result}
\end{figure}

Figure~\ref{fig:community_detection_result} compares the resulting communities
for four disease-conditioned networks with the MSigDB-only baseline. Differences
between the disease-conditioned panels and the baseline indicate how the sparse
DGIdb layer reorganizes the background functional geometry. Communities that
persist across diseases primarily reflect structure inherited from MSigDB,
whereas communities that change across diseases or contain many DGIdb-associated
genes reflect the influence of disease-specific coupling.

The disease-conditioned community structures in Figure~\ref{fig:community_detection_result} exhibit both conserved and disease-specific organization. Several large communities persist across all
diseases and the MSigDB-only baseline, indicating stable background functional
structure inherited from the shared MSigDB layer. These recurring communities
are typically enriched for broad processes such as metabolism, transcription,
cellular regulation, and vesicle-mediated transport.

At the same time, the disease-conditioned networks show substantial
reorganization relative to the MSigDB-only baseline. Neuropsychiatric diseases
display stronger enrichment for neuronal system, signal transduction,
neurotransmission, ion-channel, and GPCR-related categories, whereas Leukemia
and Breast Cancer exhibit stronger enrichment for immune-system, RNA metabolism,
translation, ribosome biogenesis, mitochondrial, extracellular-matrix, and
cell-cycle-related categories. The pie-chart distributions also show that some
communities are functionally concentrated, with enrichment dominated by a small
number of coherent categories, while others contain broader mixtures of
functional annotations. Large communities can yield many significant terms, so biological interpretation is strongest when multiple retained terms point to a common process.

\paragraph{Biological validation and hypothesis generation: }
Several category-coherent communities are consistent with known disease biology. For the neuropsychiatric diseases, enrichment for signal transduction, GPCR-related pathways, ion-channel activity, and neurotransmission is plausible. GPCR-linked signaling and downstream second-messenger pathways have been implicated in Bipolar Disorder and mechanisms of mood-stabilizer action, while GPCR and calcium-signaling dysregulation, GABAergic neurotransmission, and neurodevelopmental processes are well-established components of Schizophrenia biology \cite{tomita2013gprotein,du2004bipolar,boczek2021gpcr,schmidt2015gaba}. Similarities between Bipolar Disorder and Schizophrenia communities are also consistent with shared molecular architecture across major psychiatric disorders \cite{brainstorm2018analysis}.

For cancer-associated communities, enrichment of RNA metabolism, translation, mitochondrial translation, ribosome biogenesis, immune response, and extracellular-matrix-related categories is consistent with known cancer mechanisms. Dysregulated ribosome biogenesis and altered translational capacity are central features of tumour growth and progression, and mitochondrial translation and mitochondrial ribosomal proteins have been implicated in Breast Cancer and other malignancies \cite{hwang2024ribosome,lin2022mrp}. Immune and antigen-presentation categories are also plausible in Leukemia, where malignant blood-cell development is closely tied to immune-cell differentiation and altered hematopoietic state.

A notable signal is the recurrence of olfactory or sensory-perception-related categories. Because these terms appear across multiple diseases, they likely reflect background structure in the shared MSigDB layer. However, ectopic olfactory receptor expression has been reported in several cancers, including Breast Cancer, where specific olfactory receptors have been associated with tumour progression, invasion, and metastasis \cite{masjedi2019olfactory,li2021or5b21,kalra2020olfactory}. We therefore treat these categories as hypothesis-generating rather than as direct evidence of disease mechanism.

\begin{figure}[htbp!]
    \centering
    \includegraphics[width=0.712\linewidth]{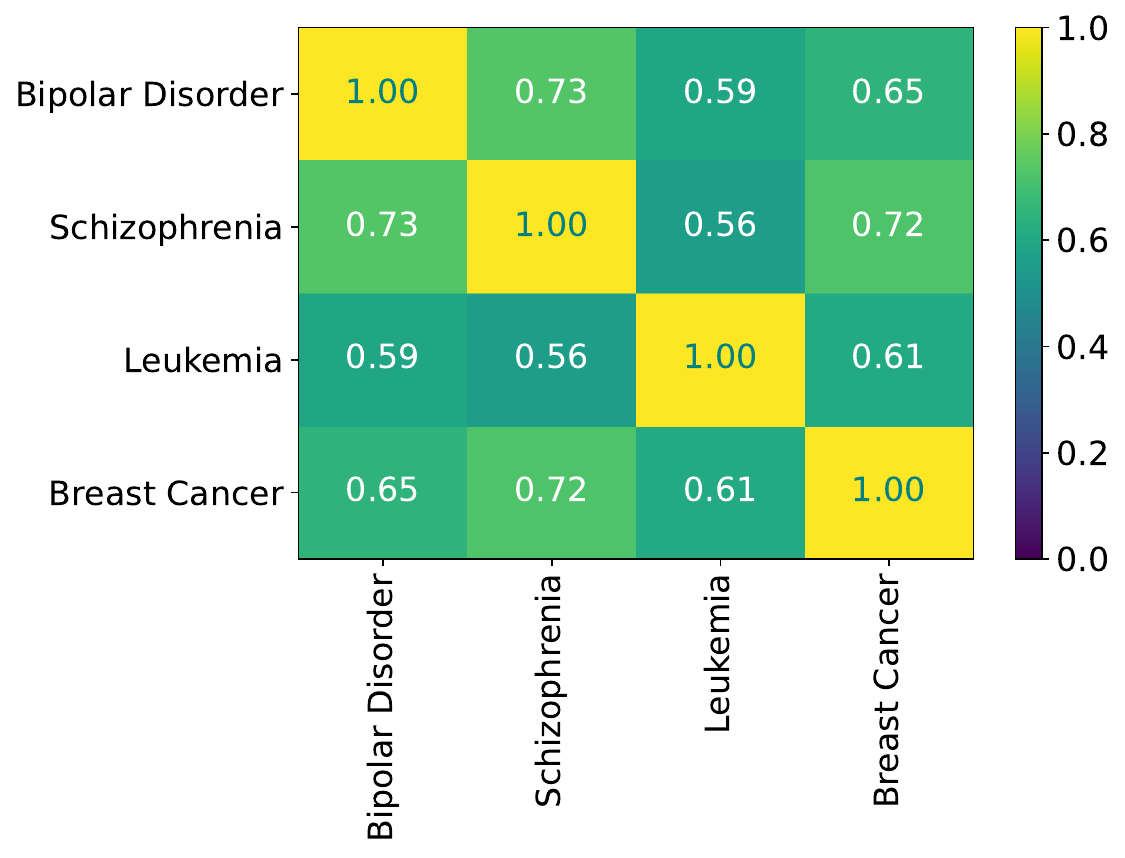}
    \caption{
        Community-based disease similarity heatmap. For each pair of diseases, selected
        communities are compared using Jaccard similarity between their gene sets.
        Disease-level similarity is computed by symmetric best-match aggregation across
        communities, as described in Appendix A7.
        The resulting values summarize similarity between diffusion-induced community
        partitions, not direct overlap between the input DGIdb gene sets. Diagonal self-comparisons are highest; Bipolar Disorder and Schizophrenia have the strongest off-diagonal similarity, while Breast Cancer shows moderate similarity with the neuropsychiatric diseases despite weaker direct DGIdb gene overlap.
        \newline
        \textbf{Alt text}: Symmetric heatmap comparing diffusion-derived community partitions across four diseases. Diagonal self-similarities equal one. Between-disease similarities range from 0.56 for schizophrenia and leukemia to 0.73 for bipolar disorder and schizophrenia. Breast cancer and schizophrenia have a similarity of 0.72.
        }
    
    \label{fig:disease_similarity_heatmap}
\end{figure}

\subsection{Disease similarity emerging from diffusion-derived communities}
\label{sec:disease_similarity}

The previous sections analyzed diffusion geometry and community structure within
individual disease-conditioned networks. We next ask whether the resulting
community organization also reveals higher-level similarity between diseases.
Importantly, our goal is not simply to compare overlap between the sparse input
DGIdb gene sets, but rather to compare the larger diffusion-induced community
decompositions that emerge after coupling to the shared MSigDB functional layer.

To quantify this effect, we compare communities across diseases using
Jaccard similarity between their gene sets and aggregate these overlaps into a
disease-level similarity score (see Appendix A7 for details).
The resulting heatmap in Figure~\ref{fig:disease_similarity_heatmap} summarizes similarity between diffusion-derived community
partitions rather than direct pharmacological overlap alone. Thus, two diseases
may appear similar either because they share DGIdb-associated genes directly or
because sparse disease-specific coupling reorganizes the broader functional
geometry in similar ways.

The largest off-diagonal similarity occurs between Bipolar Disorder and Schizophrenia, consistent with substantial evidence for shared genetic and neurobiological architecture across major psychiatric disorders, including genome-wide evidence of shared heritability and studies identifying overlapping risk loci and pathways between Bipolar Disorder and Schizophrenia \cite{brainstorm2018analysis,ruderfer2018genomic,forstner2017shared,cardno2014genetic}.
However, interestingly, Breast Cancer also exhibits moderate similarity with the neuropsychiatric
diseases, particularly Schizophrenia (with a similarity score almost as high as that between Schizophrenia and Bipolar Disorder), despite substantially weaker direct DGIdb
gene overlap (see Appendix A8 for DGIdb overlap statistics). 
Together, these results identify hypothesis-generating similarities between diffusion-derived community structures that are not fully explained by direct overlap of the sparse disease-associated input genes. They should not be interpreted as causal or database-independent evidence of biological relationships between diseases.

\paragraph{Breast Cancer--Schizophrenia link: } The Schizophrenia--Breast Cancer similarity recovered by the diffusion-derived
community comparison is consistent with recent biomedical literature, although it
should be interpreted cautiously. Population-scale and genetic studies have
reported positive phenotypic and genetic associations between schizophrenia and
breast cancer, including evidence for shared genetic contribution and positive
genome-wide genetic correlation \cite{lu2020shared,tang2023epidemiological}.
Recent clinical studies have also examined treatment-related hypotheses involving
antipsychotic exposure and prolactin, but this evidence remains mixed and
observational \cite{solmi2024antipsychotic,kern2024association}. Thus, the
similarity observed here should not be read as evidence of a direct causal
relationship. Instead, it may reflect shared genetic architecture, overlapping
drug-target or signaling modules, treatment-related annotation effects, or
structure inherited from the underlying databases. Because the analysis uses curated public resources, curation practices and preferential study of prominent genes may influence both network structure and the resulting similarities.



\subsection{Robustness analysis}
\label{sec:robustness_community_detection}
To assess robustness, we repeatedly subsample the gene set by randomly retaining
\(80\%\) of genes without replacement. For each subsample, we reconstruct the
induced diffusion-derived graph on the retained genes and rerun Leiden community
detection. For each subsample, the resulting partition is compared against the original
full-data partition after restricting both partitions to the retained genes.
Similarity is quantified using the Adjusted Rand Index (ARI) \cite{hubert1985comparing}. High ARI values indicate that the diffusion-derived community structure
is stable under moderate perturbations of the underlying gene set.
We monitor convergence of the median ARI every 50 runs and stop when the median changes by less than 0.01 for two consecutive checks, with a maximum of 1000 runs.

\begin{table}[h]
\caption{Subsampling-based robustness of the detected gene communities. The disease-conditioned networks have stability comparable to the MSigDB-only baseline.}
\label{tab:median_ARI}
\centering
\begin{tabular}{lccc}
\toprule
Disease & Median ARI & Mean ARI & Std. dev. \\
\midrule
Bipolar Disorder & 0.594 & 0.620 & 0.107 \\
Schizophrenia & 0.652 & 0.679 & 0.149 \\
Leukemia & 0.632 & 0.639 & 0.071 \\
Breast Cancer & 0.694 & 0.697 & 0.114 \\
\midrule
Averaged across diseases & 0.643 & 0.659 & 0.110 \\
MSigDB baseline & 0.582 & 0.599 & 0.087 \\
\bottomrule
\end{tabular}
\end{table}

As shown in Table~\ref{tab:median_ARI}, the median ARI values are consistently well above the level expected from random agreement. ARI values of $\sim 0.6$ indicate moderate to strong robustness of the community structure under perturbations of the gene set. 
The disease-conditioned networks exhibit ARI values that are comparable to, and
in some cases slightly higher than, the MSigDB-only baseline. This suggests that
introducing sparse disease-specific coupling alters the diffusion-derived
community structure without reducing its robustness under subsampling.
See Appendix A15 for full distributions of ARI values under sampling for all diseases.

\section{Conclusion}
\label{sec:discussion}

We introduced a diffusion-based framework for analyzing sparsely coupled
multilayer hypergraphs and applied it to disease-conditioned gene networks. The
framework couples a sparse condition-specific hypergraph layer to a dense shared
background layer and uses random walks on the coupled system to quantify how
sparse contextual information reshapes multiscale diffusion geometry and
community structure.

Disease-specific DGIdb layers, although comprising fewer than \(2\%\) of genes
in the coupled system, substantially alter diffusion distances and community
organization. This effect can be understood as structural amplification: sparse
interlayer transitions are routed through DGIdb-associated genes that we find to occupy
relatively central or bridging positions in the MSigDB-derived functional
network.

The diffusion-derived communities group genes with similar multiscale
random-walk profiles and are stable under subsampling. We find that some communities recur across diseases and reflect background organization inherited from MSigDB,
whereas others are more disease-conditioned  or show disease-relevant enrichment. Post hoc enrichment
identifies coherent biological modules validated in biomedical literature, including signaling and
neurotransmission in neuropsychiatric diseases and immune, translational, and
metabolic processes in cancer-associated diseases.

Beyond individual disease networks, the diffusion-derived communities reveal
disease-level similarities not reducible to direct DGIdb gene overlap or local
network adjacency alone. In particular, the emergence of a
Schizophrenia--Breast Cancer similarity in our data-driven approach captures recent biomedical evidence for
shared genetic or phenotypic association.

Overall, sparse contextual coupling provides a general mechanism by which small
condition-specific layers can induce interpretable, nonlocal changes in
higher-order network geometry. This framework is not specific to gene networks
and applies whenever a dense higher-order layer is coupled to sparse contextual
information. A limitation of the present study is its reliance on incomplete and
potentially biased curated databases. Although MSigDB and DGIdb aggregate evidence across many studies and resources, biases pervasive in the underlying literature can propagate into the analysis, as in other data-driven studies. High degree may reflect preferential study, genuine biological pleiotropy, or both; an arbitrary hub penalty could therefore remove biological signal together with potential bias while defining a different random walk. Diffusion proximity and the Breast Cancer--Schizophrenia comparison should consequently be interpreted as hypothesis-generating similarities in diffusion-derived organization rather than mechanistic, causal, or database-independent evidence. Future work
will incorporate additional biological layers, patient-specific data,
alternative coupling schemes, scalable approximations, and theoretical
conditions under which sparse contextual layers produce large geometric effects.

\section*{Funding}
\addcontentsline{toc}{section}{Funding}
No funding was received for this work.

\section*{Acknowledgments}
\addcontentsline{toc}{section}{Acknowledgments}
We thank Richard Huang for insightful discussions. The authors acknowledge the use of LLMs for editing, proofreading, and code-formatting assistance. No LLMs were used for defining the problem statement, developing the methodology, and interpreting the results. The authors take full responsibility for the manuscript.

\section*{Code and data availability}
\addcontentsline{toc}{section}{Code and data availability}

The code used for the analyses is available at
\url{https://github.com/Monomanae/DDBC-hypergraph}.
All datasets used in this study are publicly available from DDDB, DGIdb,
MSigDB, and HumanNet-GSP, as cited in the manuscript.

\section*{Appendix}
\label{sec:appendix}
\addcontentsline{toc}{section}{Appendix}
\setcounter{subsection}{0}
\renewcommand{\thesubsection}{A.\arabic{subsection}}
This appendix provides additional construction details, implementation choices,
and supplementary analyses for the multilayer hypergraph diffusion framework.
The main text describes the conceptual pipeline; here we give the precise
definitions used for hypergraph construction, layer coupling, aggregation of
duplicated genes, diffusion-distance computation, enrichment analysis,
community-level visualization, disease similarity, and stability assessment.

\subsection{Multilayer hypergraph construction}
\label{appendix:hypergraph_construction}

For each disease \(b\), we construct a two-layer hypergraph
\begin{equation}
\mathcal H_b=\{\mathcal H_b^{(D)},\mathcal H^{(M)}\},
\end{equation}
where \(\mathcal H_b^{(D)}\) is a disease-specific DGIdb drug--gene layer and
\(\mathcal H^{(M)}\) is a shared MSigDB functional gene-set layer. The DGIdb
layer changes with the disease, while the MSigDB layer is fixed across diseases.

\paragraph{HumanNet-GSP weights.}
Let \(\mathcal N=(V_{\mathcal N},E_{\mathcal N})\) denote the HumanNet-GSP
functional gene network. For a gene \(v\), let
\(\deg_{\mathrm{\mathcal N}}(v)\) denote its degree in HumanNet-GSP. We define
\begin{equation}
w(v)=
\begin{cases}
\deg_{\mathrm{\mathcal N}}(v), & v\in V_{\mathcal N},\\
0.01, & v\notin V_{\mathcal N}.
\end{cases}
\end{equation}

\paragraph{MSigDB layer.}
The MSigDB layer represents curated functional gene sets as hyperedges:
\begin{equation}
E^{(M)}=\{e:\ e \text{ is a gene set in MSigDB C2}\},
\qquad
V^{(M)}=\bigcup_{e\in E^{(M)}} e.
\end{equation}
The weighted MSigDB hypergraph is
\begin{equation}
\mathcal H^{(M)}=(V^{(M)},E^{(M)},w^{(M)}),
\end{equation}
where \(w^{(M)}\) is the restriction of \(w\) to \(V^{(M)}\). This layer captures
background functional organization and is shared across all diseases.

\paragraph{DGIdb disease-specific layer.}
For a disease \(b\), let \(\mathcal D_b\) denote the set of drugs associated
with \(b\) in DDDB. Each drug \(d\in\mathcal D_b\) defines
a hyperedge \(e_d\) consisting of its DGIdb target genes:
\begin{equation}
E_b^{(D)}=\{e_d:\ d\in\mathcal D_b\},
\qquad
V_b^{(D)}=\bigcup_{e\in E_b^{(D)}} e.
\end{equation}
The disease-specific drug--gene hypergraph is
\begin{equation}
\mathcal H_b^{(D)}=(V_b^{(D)},E_b^{(D)},w_b^{(D)}),
\end{equation}
where \(w_b^{(D)}\) is again inherited from the HumanNet-GSP weighting function.
This layer supplies the sparse disease-conditioned pharmacological information.

\subsection{Aggregation of layer-specific gene copies}
\label{appendix:aggregated_markov_operator}

The Markov operator \(P\) is defined on supra-nodes, but downstream analyses are
performed on unique genes. We therefore aggregate layer-specific copies after
diffusion.

Let \(G_D\) and \(G_M\) denote the gene sets in the DGIdb and MSigDB layers,
respectively. The column aggregation matrix
\begin{equation}
A_c\in \mathbb R^{(|G_D|+|G_M|)\times |G_D\cup G_M|}
\end{equation}
is defined by
\begin{equation}
(A_c)_{ij}=
\begin{cases}
1, & \text{if supra-node } i \text{ is a copy of gene } j,\\
0, & \text{otherwise.}
\end{cases}
\end{equation}
Multiplication by \(A_c\) collapses destination probabilities assigned to
multiple layer-specific copies of the same gene.

For row aggregation, let \(\pi\) denote the stationary distribution of \(P\):
\begin{equation}
\pi^\top P=\pi^\top,
\qquad
\pi_i\ge 0,
\qquad
\sum_i \pi_i=1.
\end{equation}
In implementation, \(\pi\) is computed on the largest recurrent component, or equivalently from the normalized leading left eigenvector of \(P\); when multiple stationary distributions exist, a small teleportation regularization can be added.

If a gene appears in only one layer, its row aggregation weight is \(1\). If
gene \(g\) appears in both layers with DGIdb copy \(j_D\) and MSigDB copy
\(j_M\), we define
\begin{equation}
(A_r)_{g,j_D}
=
\frac{\pi(j_D)}{\pi(j_D)+\pi(j_M)},
\qquad
(A_r)_{g,j_M}
=
\frac{\pi(j_M)}{\pi(j_D)+\pi(j_M)}.
\end{equation}
All other entries in row \(g\) are zero. The aggregated \(t\)-step diffusion
profile matrix is
\begin{equation}
\widetilde P^t=A_rP^tA_c.
\end{equation}
This matrix has one row and one column per unique gene.



\subsection{Hyperedge cardinality and one-step retention}
\label{appendix:hyperedge_retention}

The cardinality normalization in the incidence-based walk prevents a large hyperedge from receiving more total transition probability merely because it contains more genes. Larger pathways may nevertheless differ systematically in their incidence and overlap patterns. We therefore directly tested whether large MSigDB hyperedges retain more diffusion mass than arbitrary gene sets of the same cardinality.

For the coupled transition matrix $P$, we calculated its stationary distribution $\pi$ and numerically verified that its entries were nonnegative, summed to one, and satisfied $\pi P\approx\pi$ to numerical precision. For each MSigDB hyperedge $e$ containing at least two represented MSigDB-layer genes, we defined the stationary one-step retention probability
\begin{equation}
R(e)=
\frac{\sum_{i\in e}\pi_i\sum_{j\in e}P_{ij}}
{\sum_{i\in e}\pi_i}.
\end{equation}
This is the probability that a walker initialized according to the stationary distribution restricted to $e$ remains within the pathway after one transition. To remove the automatic effect of set size, we sampled 500 random subsets of MSigDB-layer genes without replacement for each hyperedge, with every random set having cardinality $|e|$. If $\mu_{|e|}$ and $\sigma_{|e|}$ denote the mean and standard deviation of their retention values, the standardized excess retention is
\begin{equation}
Z(e)=\frac{R(e)-\mu_{|e|}}{\sigma_{|e|}}.
\end{equation}
A value near zero indicates retention comparable to an arbitrary gene set of the same size, whereas a positive value indicates greater retention than expected from cardinality alone.

\begin{figure}[htbp!]
    \centering
    \includegraphics[width=0.72\linewidth]{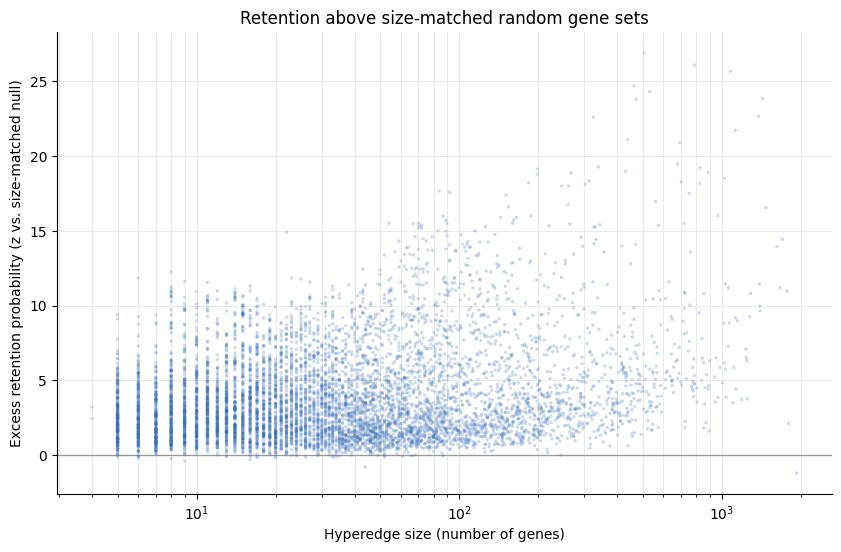}
    \caption{Standardized excess one-step retention of MSigDB hyperedges relative to 500 size-matched random gene sets, plotted against hyperedge size on a logarithmic horizontal axis. Values are broadly dispersed across sizes without a strong increasing trend. No strong correlation is observed, indicating that large hyperedges do not appear to behave as topological traps.
    \newline
    \textbf{Alt text}: Scatter plot of standardized excess one-step retention for MSigDB hyperedges versus hyperedge size on a logarithmic horizontal axis. Most points lie above the zero reference line with values broadly dispersed across sizes.}
    \label{fig:hyperedge_retention}
\end{figure}

As shown in Figure~\ref{fig:hyperedge_retention}, standardized excess retention does not increase systematically with hyperedge size. If cardinality itself created topological traps, an increasing relationship would be expected (approximately linear on the plotted $\log_{10}|e|$ scale). Instead, the broad dispersion across sizes provides no strong evidence that large hyperedges are inherently traps. A full spectral analysis of the weighted coupled operator would be interesting future work, but is beyond the present scope because it must jointly account for heterogeneous weights, hyperedge overlaps, and interlayer coupling. The size-matched retention analysis directly evaluates the proposed trapping mechanism without making broader claims about individual slow eigenmodes.

\subsection{\texorpdfstring{\(k\)}{k}-nearest-neighbor graph construction}
\label{appendix:knn_graph}

The multiscale squared diffusion distance matrix is converted to a similarity
matrix using a Gaussian kernel:
\begin{equation}
K_{ij}
=
\exp\left(-\frac{\overline D^2_{ij}}{2\sigma^2}\right),
\end{equation}
where \(\sigma\) is the median of the nonzero entries of \(\overline D\), the average diffusion distance matrix built in Section 3.5. For
each gene, we retain the \(k\) largest similarities, corresponding to the
\(k\) nearest neighbors in diffusion space. The resulting directed matrix is
symmetrized by
\begin{equation}
\widetilde K
\leftarrow
\frac{1}{2}(\widetilde K+\widetilde K^\top).
\end{equation}
The weighted undirected graph represented by \(\widetilde K\) is used for
Leiden community detection. It is an affinity graph in diffusion-profile space, not a clique expansion: its edges encode proximity under multiscale diffusion through the full multilayer hypergraph rather than co-membership in an input hyperedge.

\subsection{Community filtering, enrichment, and categorization}
\label{appendix:enrichment_and_categorization}

Let \(C\) be a community detected in the diffusion-derived \(k\)NN graph.
Communities smaller than a minimum size threshold \(s\) are removed from
downstream enrichment analysis. Within each retained community, we optionally
retain only central genes. Let \(H=\widetilde K[C]\) be the induced weighted
subgraph on \(C\). For each \(u\in C\), define its within-community weighted
degree
\begin{equation}
\text{deg}_C(u)=\sum_{v\in C} w_{uv}.
\end{equation}
Let \(z_C(u)\) be the z-score of \(\text{deg}_C(u)\) within the community. Genes
satisfying
\begin{equation}
z_C(u)\ge z
\end{equation}
are retained for enrichment analysis.

Functional enrichment analysis was performed using GSEApy's Enrichr interface against the GO Biological Process 2023, GO Molecular Function 2023, GO Cellular Component 2023, KEGG 2021 Human, and Reactome 2022 libraries \cite{fang2022gseapy,kuleshov2016enrichr,Aleksander2023geneontology,kanehisa2025kegg,kanehisa2019origin,kanehisa2000kegg,gillespie2022reactome}. Enriched terms were further grouped into higher-level functional categories using GO, KEGG, and Reactome pathway annotations. Enrichr computes a one-sided Fisher exact test for the
overlap between the community gene set and each term gene set. Resulting
\(p\)-values are adjusted using the Benjamini--Hochberg procedure. Each term has
an adjusted \(p\)-value \(p_{\mathrm{adj}}\) and overlap \(a/b\), where \(a\) is
the number of overlapping genes and \(b\) is the size of the term gene set. We
retain terms satisfying
\begin{equation}
p_{\mathrm{adj}}<\alpha,
\qquad
\frac{a}{b}>\tau.
\end{equation}
Retained terms are then mapped to broad functional categories for visualization
and interpretation. This enrichment step is post hoc: it is not used to define
the diffusion operator, the diffusion distances, or the communities.

\subsection{Community graph visualization}
\label{appendix:community_similarity_graph}

To visualize relationships among communities, we construct a community-level
graph. For communities \(C_1,\dots,C_n\) in a weighted graph with edge weights
\(w(u,v)\), define the mean inter-community similarity
\begin{equation}
S_{ij}
=
\frac{1}{|C_i||C_j|}
\sum_{u\in C_i}
\sum_{v\in C_j}
w(u,v),
\end{equation}
where \(w(u,v)=0\) if no edge is present. The visualization contains one node per
community and an edge between \(C_i\) and \(C_j\) if \(S_{ij}\) exceeds a small
threshold. Edge weights are normalized by the maximum retained edge weight.

\subsection{Disease similarity from community overlap}
\label{appendix:disease_similarity_jaccard}

We quantify disease-level similarity by comparing the gene composition of
selected communities. For disease \(d\), let
\begin{equation}
\mathcal C^{(d)}
=
\{C^{(d)}_1,\dots,C^{(d)}_{k_d}\}
\end{equation}
be its selected communities. For two diseases \(d_1\) and \(d_2\), define the
community-overlap matrix
\begin{equation}
J^{(d_1,d_2)}_{ij}
=
\frac{|C^{(d_1)}_i\cap C^{(d_2)}_j|}
{|C^{(d_1)}_i\cup C^{(d_2)}_j|}.
\end{equation}
Disease-level similarity is computed by symmetric best-match aggregation:
\begin{equation}
S(d_1,d_2)
=
\frac{
\sum_i \max_j J^{(d_1,d_2)}_{ij}
+
\sum_j \max_i J^{(d_1,d_2)}_{ij}
}{
k_{d_1}+k_{d_2}
}.
\end{equation}
This score is high when communities in each disease have close counterparts in
the other disease. It should be interpreted as similarity between
diffusion-induced community decompositions, not as a direct measure of causal
disease relatedness.

\subsection{DGIdb gene-set overlap across diseases}
\label{appendix:dgidb_gene_overlap}

\begin{figure}[H]
    \centering
    \includegraphics[width=0.58\textwidth]{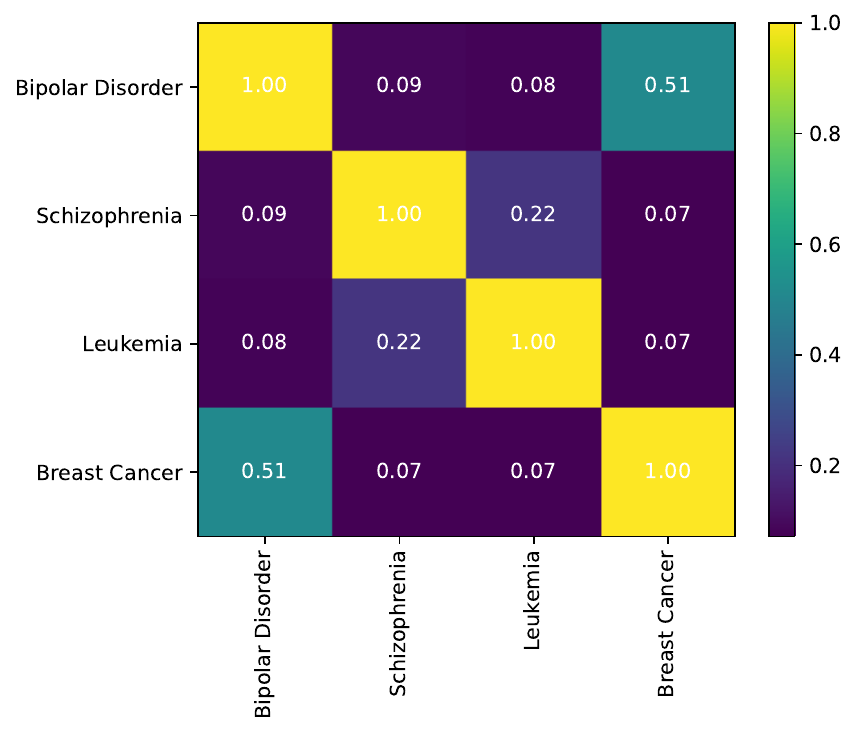}
    \caption{Jaccard overlap of disease-specific DGIdb gene sets for Bipolar
    Disorder, Schizophrenia, Leukemia, and Breast Cancer. Diagonal entries equal one and represent self-comparisons; off-diagonal entries quantify direct target-gene overlap between diseases before diffusion and show that overlap varies across disease pairs.
    \newline
    \textbf{Alt text}: Symmetric heatmap of direct Jaccard overlap between disease-specific DGIdb gene sets. Diagonal values equal one. Bipolar disorder and breast cancer have the largest overlap, 0.51, followed by schizophrenia and leukemia, 0.22. All remaining pairs have overlaps below 0.1.}
    \label{fig:gene_overlap}
\end{figure}

The overlap between disease-specific DGIdb gene sets is summarized in
Figure~\ref{fig:gene_overlap}. This comparison is included to distinguish
similarities already present in the sparse input layer from similarities emerging
after coupling to the MSigDB functional layer. For example, a high overlap in
Figure~\ref{fig:gene_overlap} indicates that two diseases share many DGIdb
target genes before diffusion. In contrast, the community-based similarity in
Figure 5 compares the resulting
diffusion-derived community partitions. Comparing the two heatmaps therefore
helps separate input-level similarity from geometry-induced similarity.

\subsection{Diffusion distance distributions for different interlayer transition probability}
\label{appendix:diffusion_distance_distribution}

\begin{figure}[p]
\centering
\vspace*{-7.5em}
\begin{subfigure}{\linewidth}
\centering
\includegraphics[width=0.85\linewidth]{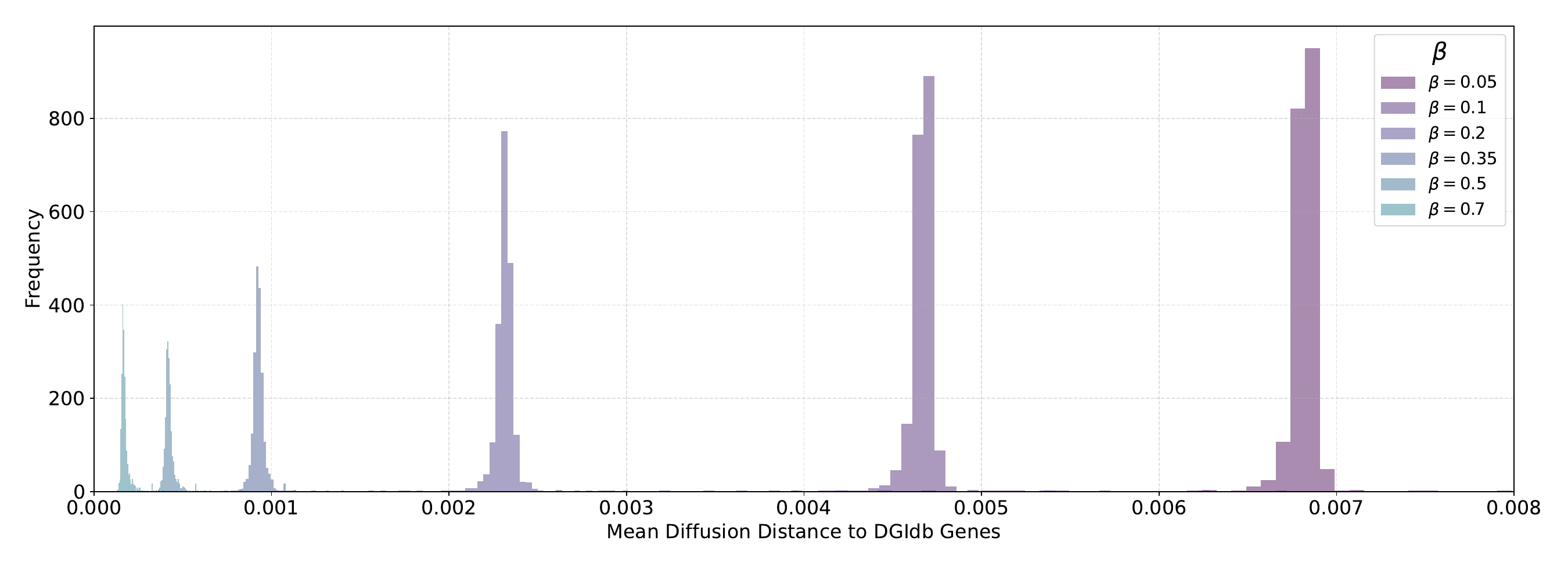}
\caption{Bipolar Disorder}
\end{subfigure}

\begin{subfigure}{\linewidth}
\centering
\includegraphics[width=0.85\linewidth]{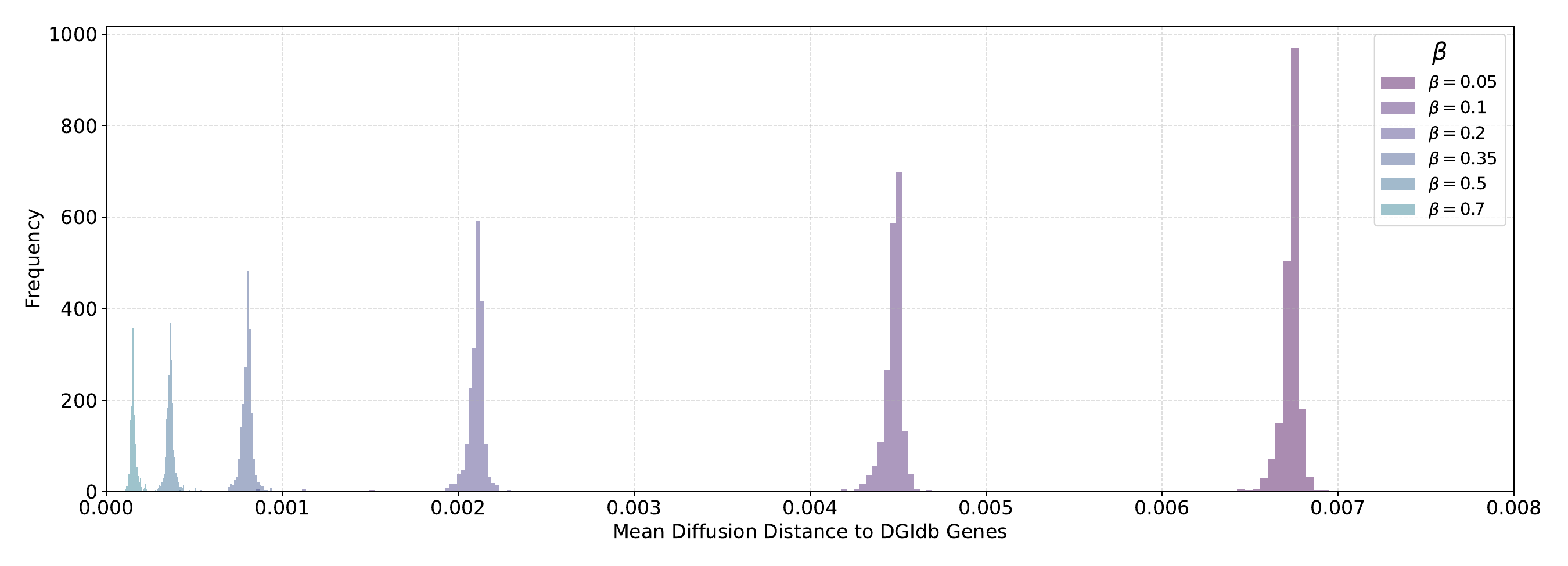}
\caption{Schizophrenia}
\end{subfigure}

\begin{subfigure}{\linewidth}
\centering
\includegraphics[width=0.85\linewidth]{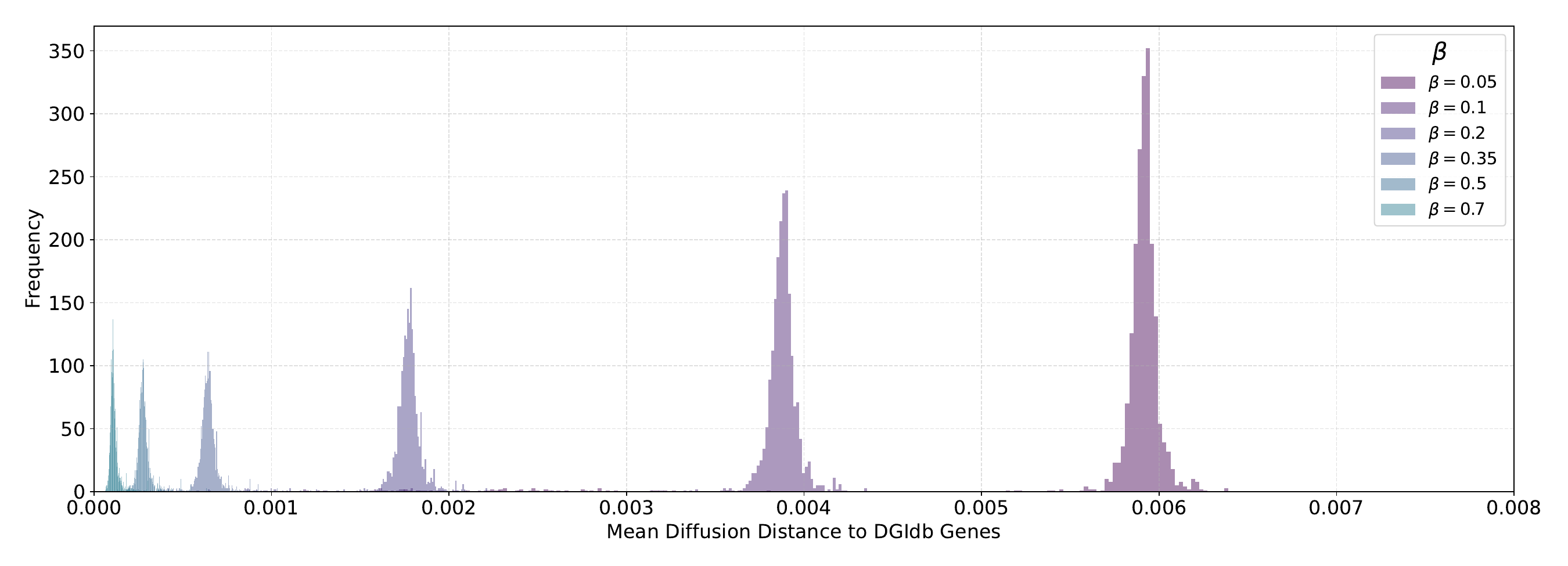}
\caption{Leukemia}
\end{subfigure}

\begin{subfigure}{\linewidth}
\centering
\includegraphics[width=0.85\linewidth]{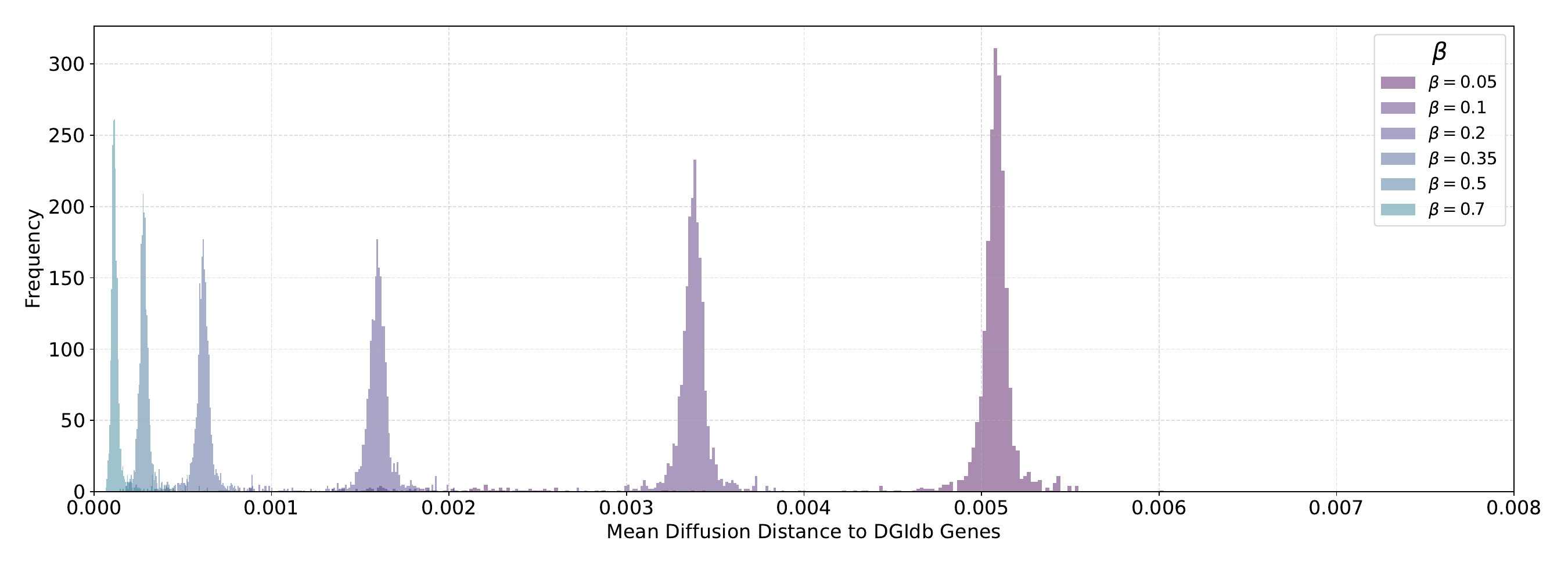}
\caption{Breast Cancer}
\end{subfigure}

    \caption{Distributions of the average squared diffusion distance from sampled genes to DGIdb genes, using the same samples as in Figure 3, for $\beta \in \{0.05, 0.1, 0.2, 0.35, 0.5, 0.7\}$, for the four diseases. In each panel, increasing $\beta$ shifts the distribution toward smaller distances and generally makes it more concentrated. The horizontal axis is cropped to \(x = 0.008\) for visualization, omitting a few outliers.
    \newline
    \textbf{Alt text}: Four histogram panels show average squared diffusion distances from sampled genes to DGIdb genes for bipolar disorder, schizophrenia, leukemia, and breast cancer. In every panel, increasing interlayer transition probability shifts the distributions toward smaller distances and generally makes them more concentrated.}

    \label{fig:all_to_dgidb_avg_diff_dist_distribution_comparison}
\end{figure}

Figure~\ref{fig:all_to_dgidb_avg_diff_dist_distribution_comparison}  shows the distributions of average squared diffusion distance to DGIdb genes for several values of $\beta$, across all four diseases. As \(\beta\) increases, the distributions shift systematically to the left,
showing that sampled genes become closer on average to DGIdb-associated genes in
diffusion space. The distributions also become more concentrated, suggesting
that the effect is broad across the sampled gene set rather than being driven
only by a small subset of strongly affected genes.

\subsection{Additional information for diffusion-derived communities}
See Table~\ref{tab:comm_stat} for more information about the diffusion-derived communities for Bipolar Disorder, Schizophrenia, Leukemia, and Breast Cancer.

\begin{longtable}{ccccc}
\caption{Community-level summary for the Bipolar Disorder, Schizophrenia, Leukemia, and Breast Cancer diffusion clusterings. ``Genes involved'' denotes the percentage of selected community genes appearing in at least one retained enriched term. ``Selected genes'' denotes the number of genes retained after the selection mechanism outlined in Appendix~\ref{appendix:enrichment_and_categorization}. ``DGIdb genes'' denotes the number of genes in the community that appear in the DGIdb network. ``Terms'' denotes the number of enriched terms in that community after filtering described in Appendix~\ref{appendix:enrichment_and_categorization}.}
\label{tab:comm_stat}\\

\toprule
Community & Genes involved (\%) & Selected genes & DGIdb genes & Terms \\
\midrule
\endfirsthead

\toprule
Community & Genes involved (\%) & Selected genes & DGIdb genes & Terms \\
\midrule
\endhead

\midrule
\multicolumn{5}{r}{Continued on next page} \\
\endfoot

\bottomrule
\endlastfoot

\multicolumn{5}{l}{\textbf{Bipolar Disorder}} \\
\midrule
0  & 0.00  & 1008 & 0   & 0   \\
1  & 57.13 & 1031 & 0   & 109 \\
2  & 32.67 & 949  & 0   & 24  \\
3  & 51.25 & 1162 & 0   & 203 \\
4  & 80.93 & 949  & 0   & 137 \\
5  & 95.58 & 837  & 0   & 654 \\
6  & 68.33 & 502  & 0   & 235 \\
7  & 52.74 & 565  & 0   & 49  \\
8  & 53.51 & 556  & 0   & 136 \\
9  & 0.00  & 340  & 0   & 0   \\
10 & 100.00 & 299 & 0   & 9   \\
11 & 44.07 & 119  & 266 & 41  \\

\midrule
\multicolumn{5}{l}{\textbf{Schizophrenia}} \\
\midrule
0  & 0.00   & 863  & 0   & 0   \\
1  & 58.30  & 1000 & 0   & 109 \\
2  & 34.91  & 951  & 0   & 24  \\
3  & 51.35  & 1152 & 0   & 203 \\
4  & 94.89  & 979  & 0   & 137 \\
5  & 93.37  & 905  & 0   & 654 \\
6  & 37.31  & 859  & 0   & 235 \\
7  & 52.74  & 548  & 0   & 49  \\
8  & 61.15  & 417  & 0   & 136 \\
9  & 0.00   & 336  & 0   & 0   \\
10 & 100.00 & 300  & 0   & 9   \\
11 & 48.03  & 127  & 249 & 41  \\

\midrule
\multicolumn{5}{l}{\textbf{Leukemia}} \\
\midrule
0  & 32.65  & 974  & 0   & 0   \\
1  & 76.81  & 1092 & 0   & 109 \\
2  & 2.04   & 884  & 0   & 24  \\
3  & 40.37  & 929  & 0   & 203 \\
4  & 52.60  & 1060 & 0   & 137 \\
5  & 95.23  & 985  & 0   & 654 \\
6  & 69.89  & 837  & 0   & 235 \\
7  & 54.88  & 512  & 0   & 49  \\
8  & 55.09  & 453  & 0   & 136 \\
9  & 0.00   & 342  & 0   & 0   \\
10 & 100.00 & 300  & 0   & 9   \\
11 & 14.63  & 123  & 198 & 41  \\

\midrule
\multicolumn{5}{l}{\textbf{Breast Cancer}} \\
\midrule
0  & 30.06  & 1018 & 0   & 0   \\
1  & 23.74  & 872  & 0   & 109 \\
2  & 62.21  & 942  & 0   & 24  \\
3  & 51.04  & 1151 & 0   & 203 \\
4  & 87.21  & 993  & 0   & 137 \\
5  & 95.65  & 966  & 0   & 654 \\
6  & 35.28  & 860  & 0   & 235 \\
7  & 56.35  & 543  & 0   & 49  \\
8  & 60.49  & 406  & 0   & 136 \\
9  & 0.00   & 342  & 0   & 0   \\
10 & 100.00 & 298  & 0   & 9   \\
11 & 33.88  & 183  & 377 & 41  \\

\end{longtable}

\subsection{Comparison of diffusion-derived communities for diseases against MSigDB-only baseline}
\begin{figure}[!t]
\centering
\begin{subfigure}{0.45\linewidth}
\centering
\includegraphics[width=\linewidth]{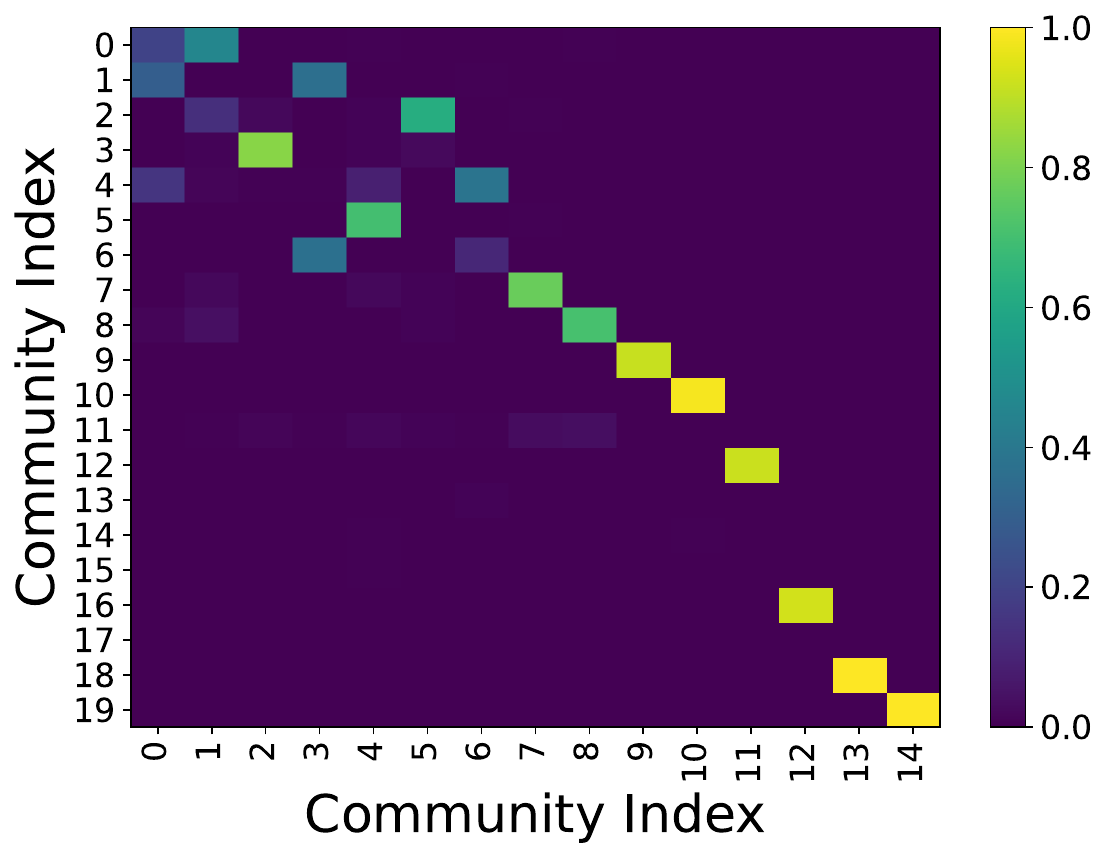}
\caption{Bipolar Disorder}
\end{subfigure}\hfill
\begin{subfigure}{0.45\linewidth}
\centering
\includegraphics[width=\linewidth]{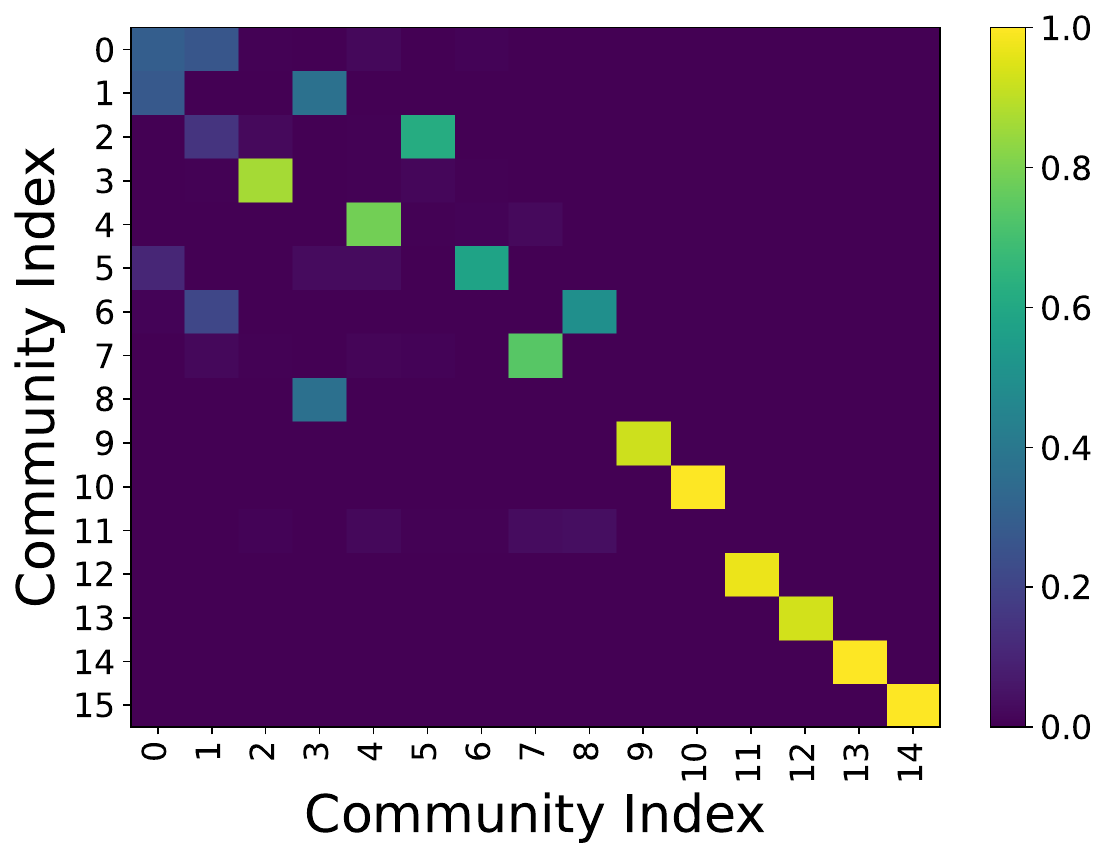}
\caption{Schizophrenia}
\end{subfigure}

\vspace{0.5em}

\begin{subfigure}{0.45\linewidth}
\centering
\includegraphics[width=\linewidth]{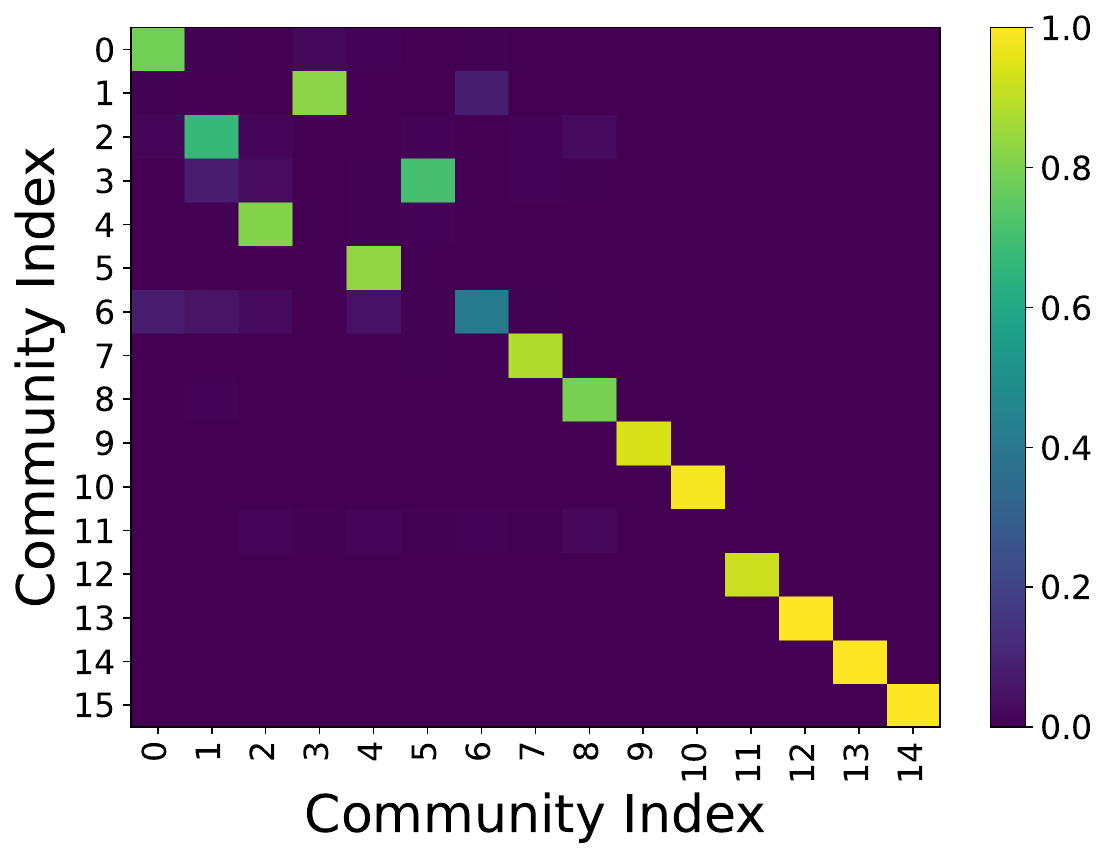}
\caption{Leukemia}
\end{subfigure}\hfill
\begin{subfigure}{0.45\linewidth}
\centering
\includegraphics[width=\linewidth]{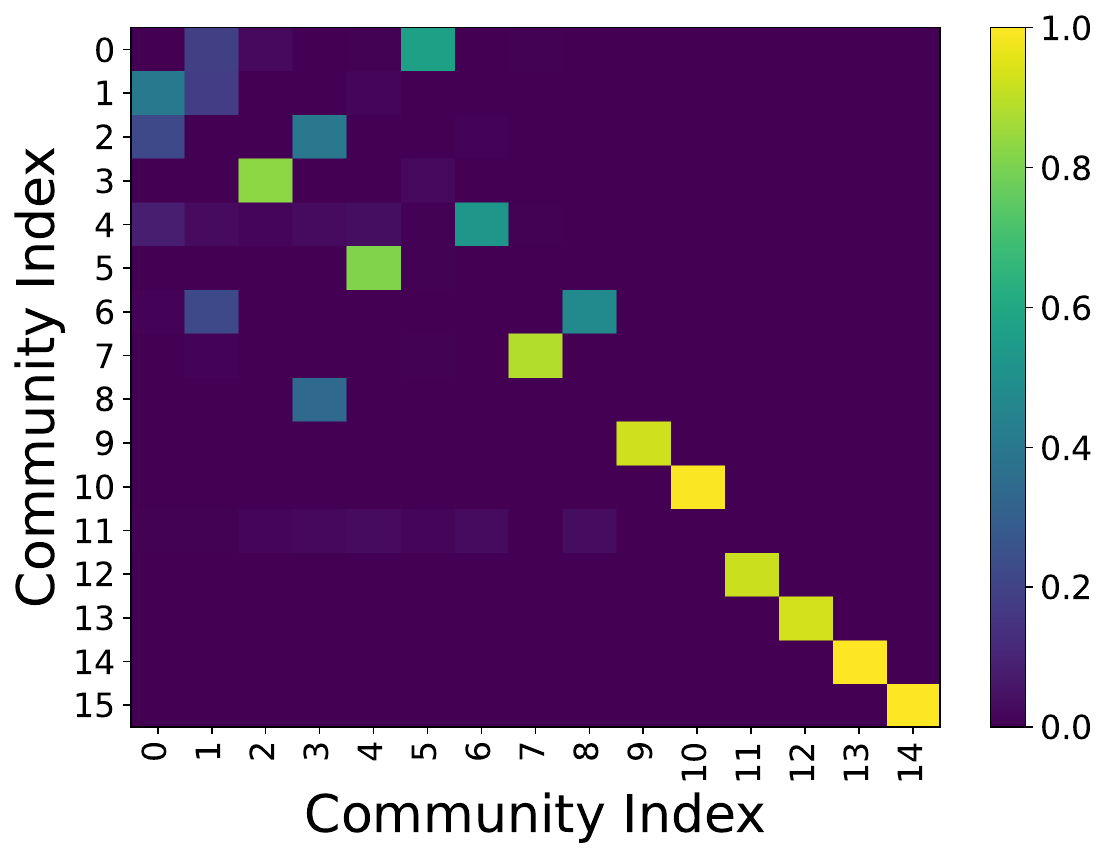}
\caption{Breast Cancer}
\end{subfigure}

\caption{Comparison of the communities of Bipolar Disorder, Schizophrenia, Leukemia, and Breast Cancer with the MSigDB-only baseline. Rows and columns represent disease-conditioned and baseline communities, respectively, and color intensity gives pairwise Jaccard similarity. Similarity is distributed across multiple cells rather than forming a one-to-one diagonal correspondence, indicating that coupling retains some baseline structure while shifting, splitting, or merging communities.
\newline
\textbf{Alt text}: Four heatmaps compare communities for bipolar disorder, schizophrenia, leukemia, and breast cancer with the MSigDB-only baseline using Jaccard similarity. Bright cells form a partial diagonal, with scattered off-diagonal matches showing overlap with multiple baseline communities.
}

\label{fig:comparison_to_baseline_jaccard}
\end{figure}

Figure~\ref{fig:comparison_to_baseline_jaccard} compares the diffusion-derived communities of Bipolar Disorder, Schizophrenia, Leukemia, and Breast Cancer  to the MSigDB-only baseline. Nonzero overlaps show that coupled communities retain substantial information from the underlying functional layer. However, the absence of a purely diagonal or one-to-one correspondence indicates that adding the DGIdb layer does not merely reproduce the MSigDB-only partition. Instead, sparse coupling reorganizes the baseline geometry by shifting, splitting, or merging functional communities. 

\subsection{Category-level enrichment scores}
\label{appendix:category_enrichment_score}

For each community \(c\) and functional category \(g\), let
\(p_1,\dots,p_{n_{c,g}}\) be the adjusted \(p\)-values of retained enriched
terms assigned to category \(g\). We summarize category-level enrichment using
Fisher's method for combining independent p-values \cite{fisher1932statistical}:
\begin{equation}
F_{c,g}
=
-2\sum_{i=1}^{n_{c,g}}\log p_i.
\end{equation}
Under the standard independence approximation, \(F_{c,g}\) is compared to a
\(\chi^2\) distribution with \(2n_{c,g}\) degrees of freedom, yielding
\(p^{\mathrm{Fisher}}_{c,g}\). We convert this value to an enrichment score
\begin{equation}
E_{c,g}
=
-\log_{10}(p^{\mathrm{Fisher}}_{c,g}).
\end{equation}
Because communities differ in size and total enrichment strength, we normalize
within each community:
\begin{equation}
Z_{c,g}
=
\frac{E_{c,g}-\mu_c}{\sigma_c},
\end{equation}
where \(\mu_c\) and \(\sigma_c\) are the mean and standard deviation of
\(E_{c,g}\) over categories for community \(c\).

This score is intended as a visualization and ranking device rather than as a
formal independent hypothesis test across categories. Terms within a category
are often correlated, so the Fisher-combined score should be interpreted as a
compact summary of category-level signal.

\subsection{Community case studies}
\label{appendix:case_studies}

The heatmaps in Figure~\ref{fig:category_gross_overlap_heatmap} summarize the
biological content of selected diffusion-derived communities at the category
level. Columns correspond to selected communities, rows correspond to broad
functional categories, color intensity shows the normalized category enrichment
score \(Z_{c,g}\) introduced in Appendix~\ref{appendix:category_enrichment_score}, and the number in each cell gives the number of retained
enriched terms assigned to that category.

The heatmaps show that many selected communities have enrichment concentrated in
a small number of categories, supporting the interpretation that these
communities correspond to coherent functional programs rather than arbitrary
graph partitions. The degree of interpretability varies: some communities have
sharply concentrated profiles, whereas others contain broad mixtures of
categories or relatively few retained terms. We therefore interpret the
communities primarily at the category level and treat isolated terms cautiously.

For neuropsychiatric diseases, several selected communities are dominated by
signal transduction, neurotransmission, ion-channel activity, and regulatory
categories. These patterns are consistent with known biology of Bipolar Disorder
and Schizophrenia. GPCR-linked signaling and downstream second-messenger
pathways have been implicated in Bipolar Disorder and mechanisms of
mood-stabilizer action, while GPCR and calcium-signaling dysregulation,
GABAergic neurotransmission, and neurodevelopmental processes are
well-established components of Schizophrenia biology
\cite{tomita2013gprotein,du2004bipolar,boczek2021gpcr,schmidt2015gaba}.
Similar high-level structure across Bipolar Disorder and Schizophrenia is also
consistent with shared molecular architecture across major psychiatric disorders
\cite{brainstorm2018analysis}.

For cancer-associated diseases, selected communities show enrichment for RNA
metabolism, translation, mitochondrial translation, ribosome biogenesis, immune
response, antigen presentation, and extracellular-matrix or cell-organization
processes. These categories are biologically plausible in cancer contexts.
Dysregulated ribosome biogenesis and altered translational capacity are central
features of tumour growth and progression, and mitochondrial translation and
mitochondrial ribosomal proteins have been implicated in Breast Cancer and other
malignancies \cite{hwang2024ribosome,lin2022mrp}. Immune and
antigen-presentation categories are also plausible in Leukemia, where malignant
blood-cell development is closely linked to immune-cell differentiation and
altered hematopoietic state.

A notable pattern is the appearance of sensory-perception or olfactory-related
categories in some selected communities. Because these categories can appear
across multiple diseases and may occur in communities with few DGIdb genes, they
should be interpreted cautiously: in the present analysis they may largely
reflect background structure inherited from the shared MSigDB layer. However,
ectopic olfactory receptor expression has been reported in several cancers,
including Breast Cancer, and specific olfactory receptors have been associated
with tumour progression, invasion, and metastasis
\cite{masjedi2019olfactory,li2021or5b21,kalra2020olfactory}. We therefore view
these categories as hypothesis-generating signals rather than direct evidence of
disease-specific mechanism.

\begin{figure}[H]
    \centering
    \begin{subfigure}{\textwidth}
        \centering
        \includegraphics[width=\textwidth,height=0.86\textheight,keepaspectratio]{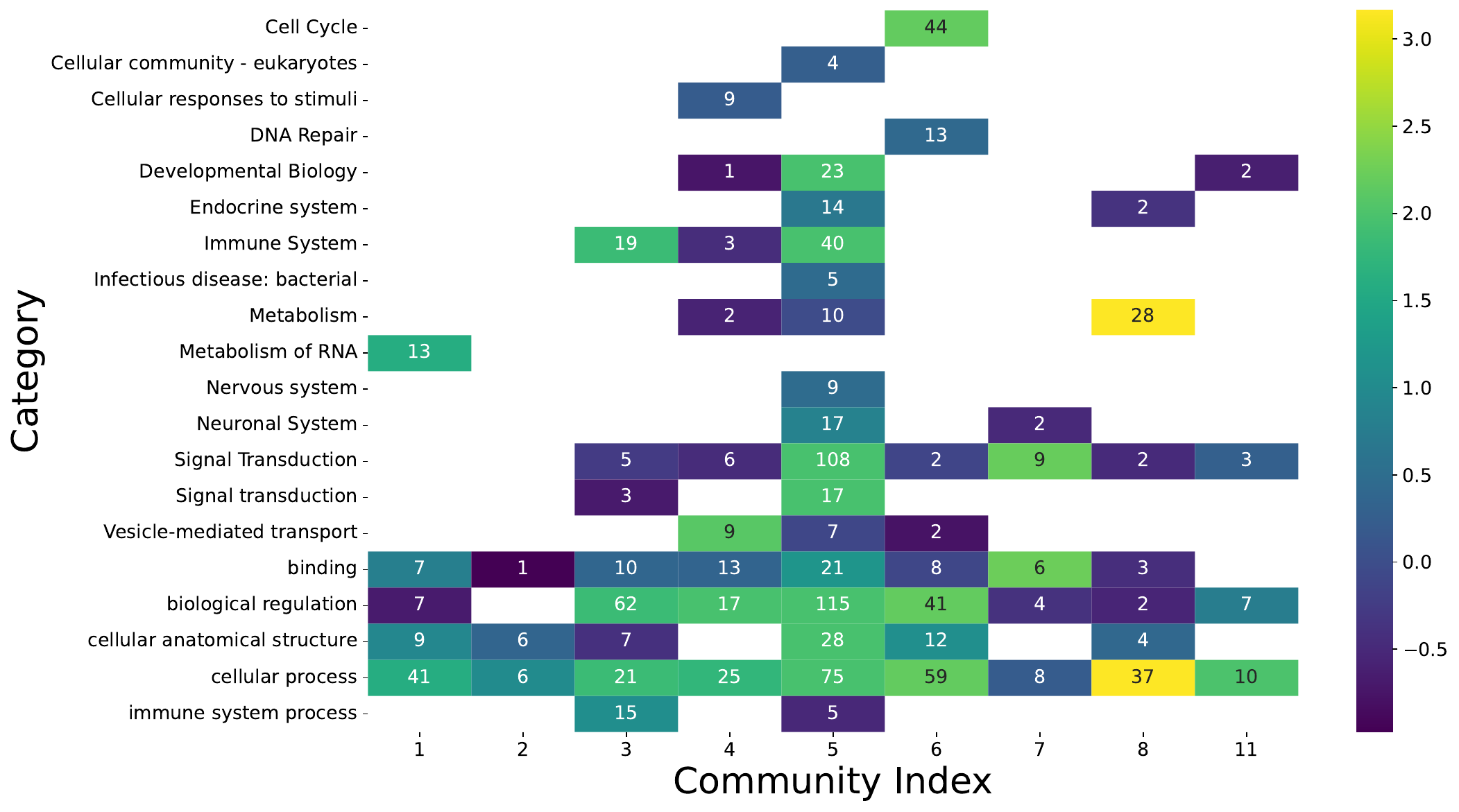}
        \caption{Bipolar Disorder}
    \end{subfigure}
\end{figure}

\begin{figure}[p]\ContinuedFloat
    \centering
    \begin{subfigure}{\textwidth}
        \centering
        \includegraphics[width=\textwidth,height=0.86\textheight,keepaspectratio]{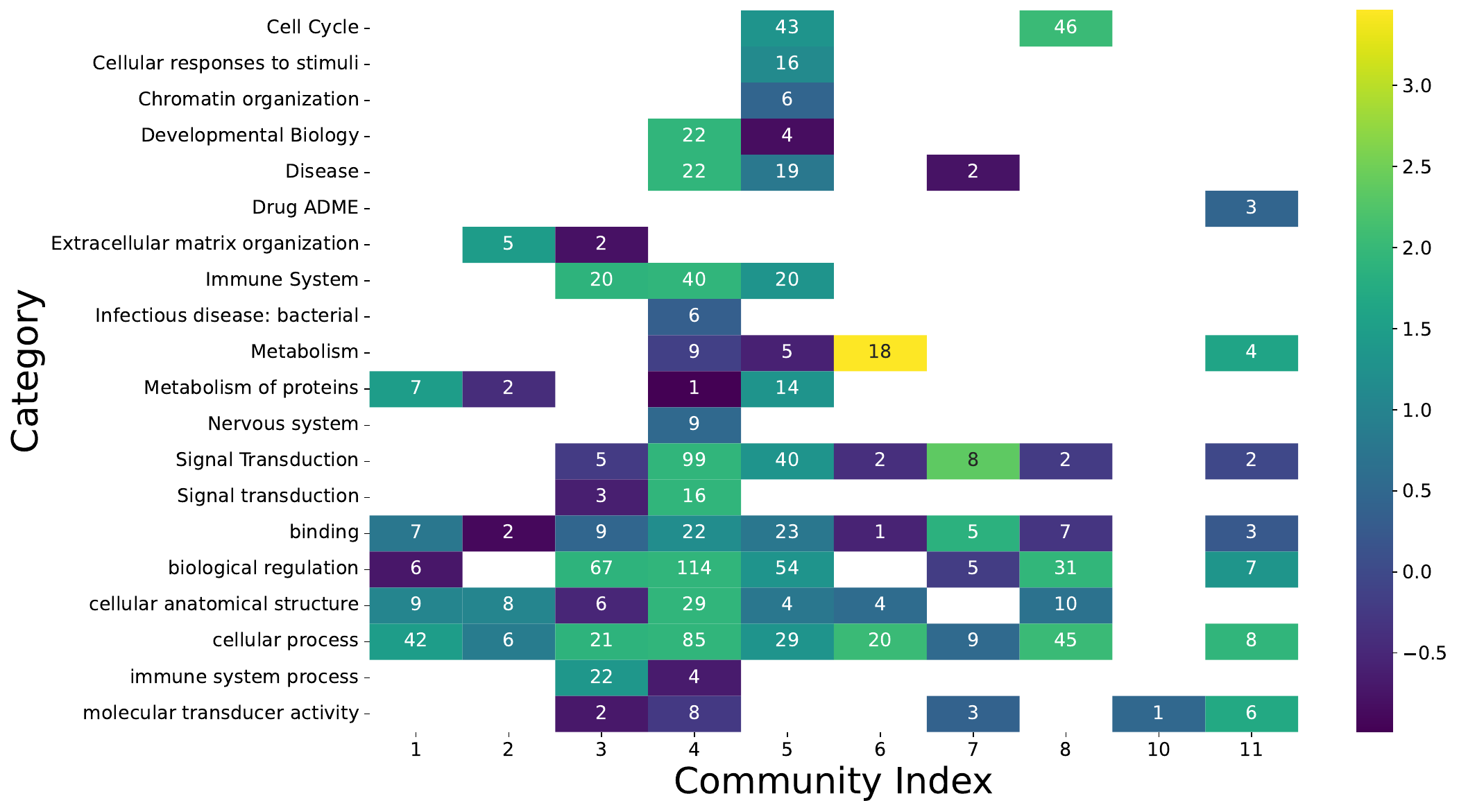}
        \caption{Schizophrenia}
    \end{subfigure}
\end{figure}

\begin{figure}[p]\ContinuedFloat
    \centering
    \begin{subfigure}{\textwidth}
        \centering
        \includegraphics[width=\textwidth,height=0.86\textheight,keepaspectratio]{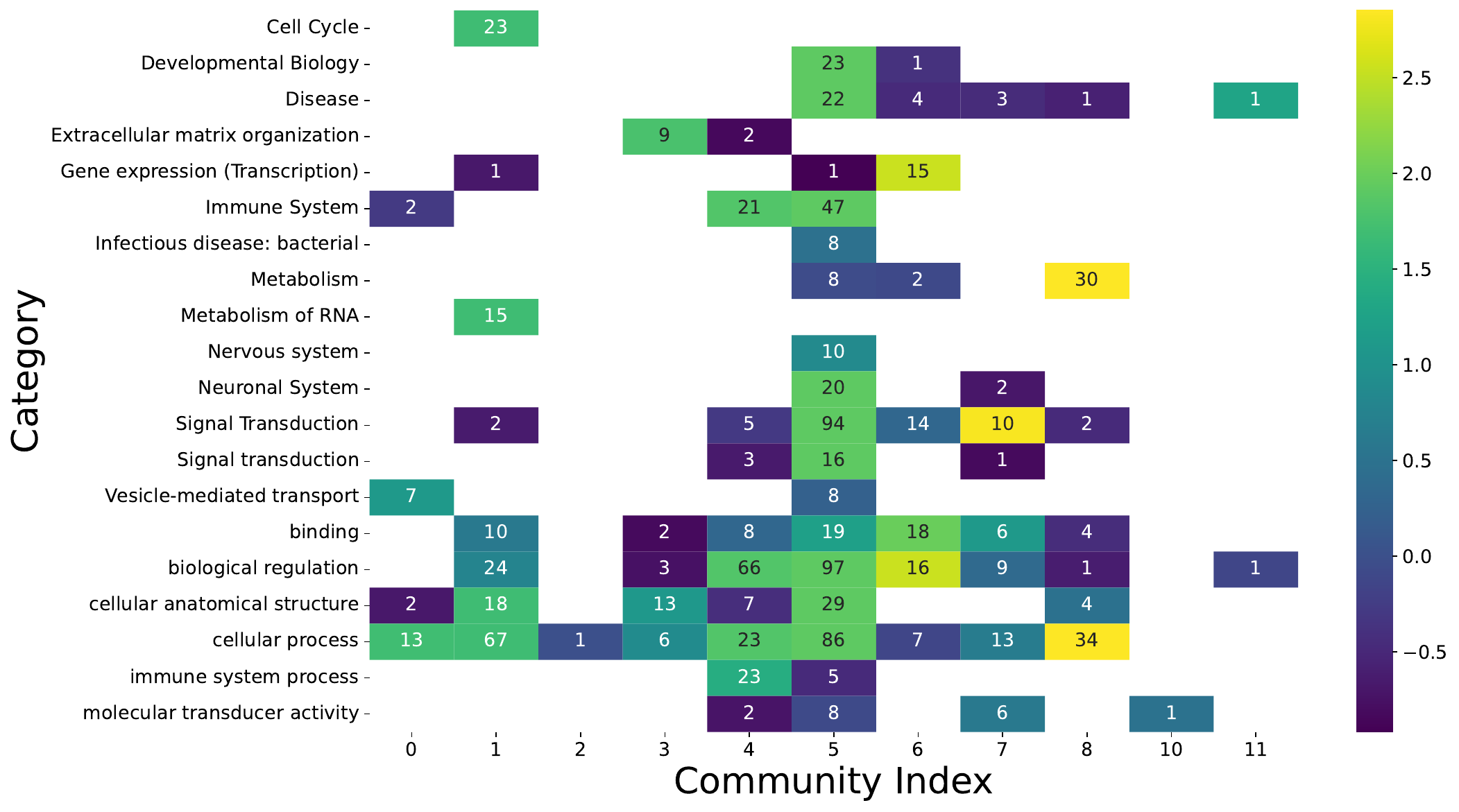}
        \caption{Leukemia}
    \end{subfigure}
\end{figure}

\begin{figure}[H]\ContinuedFloat
    \centering
    \begin{subfigure}{\textwidth}
        \centering
        \includegraphics[width=\textwidth,height=0.86\textheight,keepaspectratio]{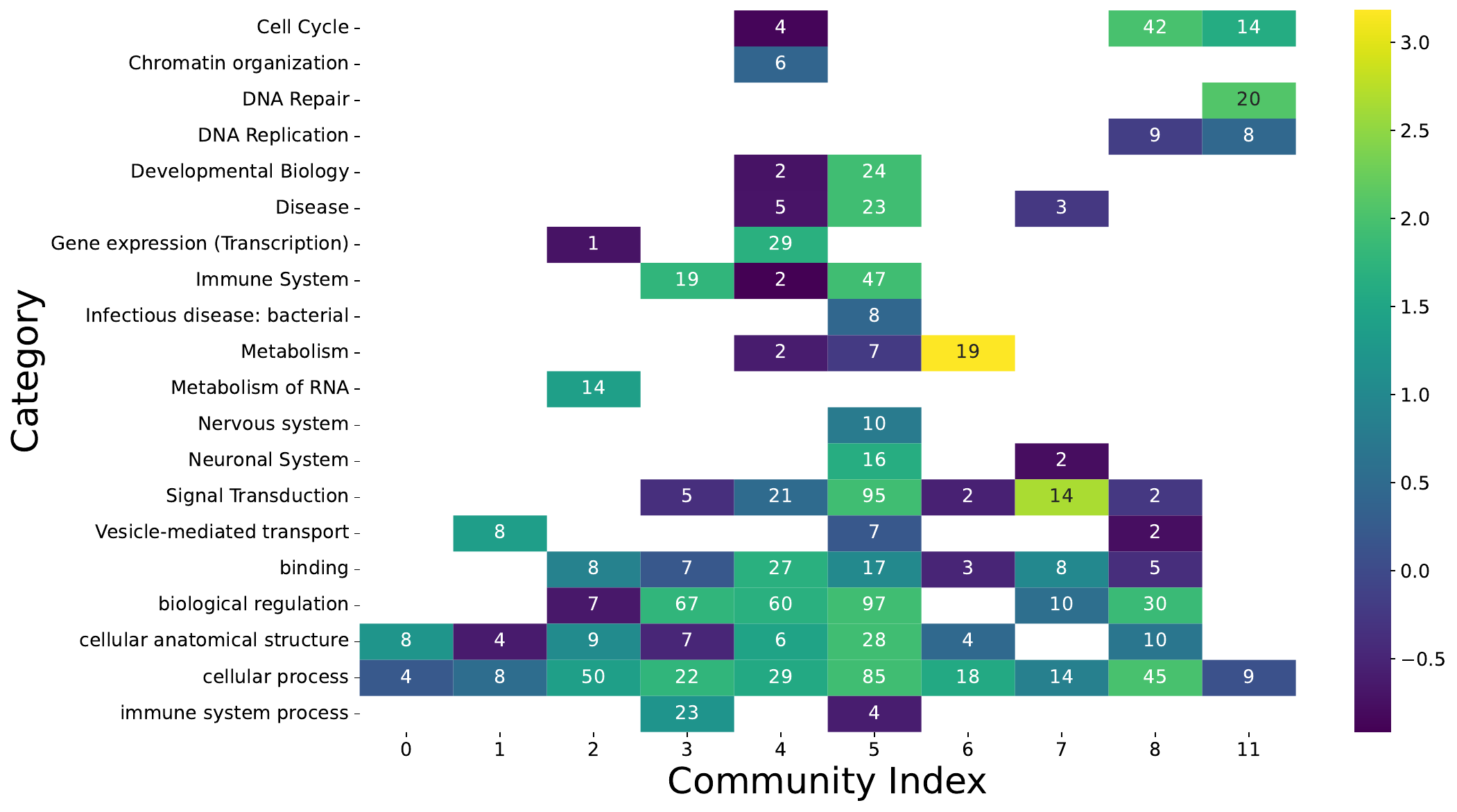}
        \caption{Breast Cancer}
    \end{subfigure}

    \caption{Category enrichment score heatmaps for selected communities in
    Bipolar Disorder, Schizophrenia, Leukemia, and Breast Cancer. Columns
    correspond to selected diffusion-derived communities and rows correspond to
    functional categories. Color intensity indicates the z-score-normalized
    category enrichment score defined in
    Appendix~\ref{appendix:category_enrichment_score}. The number displayed in
    each cell is the number of retained enriched terms in that category. Only the top 20 categories by average normalized category enrichment score across all communities are shown for better visualization. The heterogeneous patterns show that signaling, neurotransmission, developmental, metabolic, translational, immune, and cell-organization categories concentrate in different communities and diseases.
    \newline
    \textbf{Alt text}: Four heatmaps with community indices along the horizontal axes and functional categories along the vertical axes. Numbered cells range from purple through green to yellow. White gaps separate uneven patches of colored cells; some rows contain several adjacent cells, while others contain isolated cells. Vertical color scales accompany the heatmaps.}
    \label{fig:category_gross_overlap_heatmap}
\end{figure}

\begingroup
\raggedbottom

\subsection{Representative enriched terms for selected communities}
\label{appendix:representative_terms}

To complement the category-level heatmaps, representative enriched terms can be
reported for selected case-study communities. When selecting case-study communities, we prioritize disease-specific communities, i.e. communities with low Jaccard overlap with communities from other diseases. We use the following
two-stage selection procedure. First, for each community, select the two or three
dominant categories according to the category enrichment score in
Appendix~\ref{appendix:category_enrichment_score}. Second, within each selected
category, rank retained terms by
\begin{equation}
-\log_{10}(p_{\mathrm{adj}})+\log(1+a),
\end{equation}
where \(a\) is the number of overlapping genes in the term and $p_{\mathrm{adj}}$ is the adjusted \(p\)-value of the term (see Appendix~\ref{appendix:enrichment_and_categorization}). 
This favors terms that are both statistically significant and supported by nontrivial overlap, while avoiding long lists of
nearly duplicate annotations. See Table~\ref{tab:representative_case_study_terms} for the list of representative enriched terms.

\begin{table}[htbp!]
\caption{Representative enriched terms for selected case-study communities. Full enriched-term outputs are available at \href{https://github.com/Monomanae/DDBC-hypergraph/tree/main/enriched_terms}{GitHub}.} 
\label{tab:representative_case_study_terms}
\centering
\footnotesize 
\setlength{\tabcolsep}{3pt}
\renewcommand{\arraystretch}{1.08}
\begin{tabular}{p{2.3cm} p{1.5cm} p{5.3cm} c c}
\toprule
Disease & Community & Representative enriched term & Adj. $p$-value & Overlap \\
\midrule
Bipolar Disorder & C11 & G Protein-Coupled Serotonin Receptor Activity & 9.07e-12 & 8/21 \\
Bipolar Disorder & C11 & High Voltage-Gated Calcium Channel Activity & 9.97e-13 & 9/23 \\
Schizophrenia & C5 & PTEN Regulation & 2.92e-53 & 69/139 \\
Schizophrenia & C5 & PIP3 Activates AKT Signaling & 2.11e-32 & 70/268 \\
Breast Cancer & C0 & Elastic Fibre Formation & 3.05e-07 & 14/39 \\
Breast Cancer & C11 & PCNA-Dependent Long Patch Base Excision Repair & 5.08e-12 & 9/21 \\
Leukemia & C6 & Transcriptional Regulation By TP53 & 1.32e-21 & 64/354 \\
Leukemia & C6 & NOTCH1 Intracellular Domain Regulates Transcription & 7.72e-14 & 20/48 \\
\bottomrule
\end{tabular}
\end{table}

These representative terms are intended to make the category-level
interpretation auditable. They should not be interpreted as inputs to the
method: community detection is performed entirely from diffusion geometry, and
enrichment is used only afterward for biological interpretation.

\subsection{Subsampling stability distributions}
\label{appendix:ari_distributions}

Figure~\ref{fig:ari_stabilities} shows the distributions of Adjusted Rand Index
scores obtained in the subsampling-based stability analysis. Each score compares
the partition obtained from an \(80\%\) induced subsample to the restriction of
the full-data partition to that same subset of genes. The number of runs is
chosen adaptively as described in Section 4.6.

\begin{figure}[htbp!]
\centering
\begin{subfigure}{0.50\linewidth}
\centering
\includegraphics[width=\linewidth]{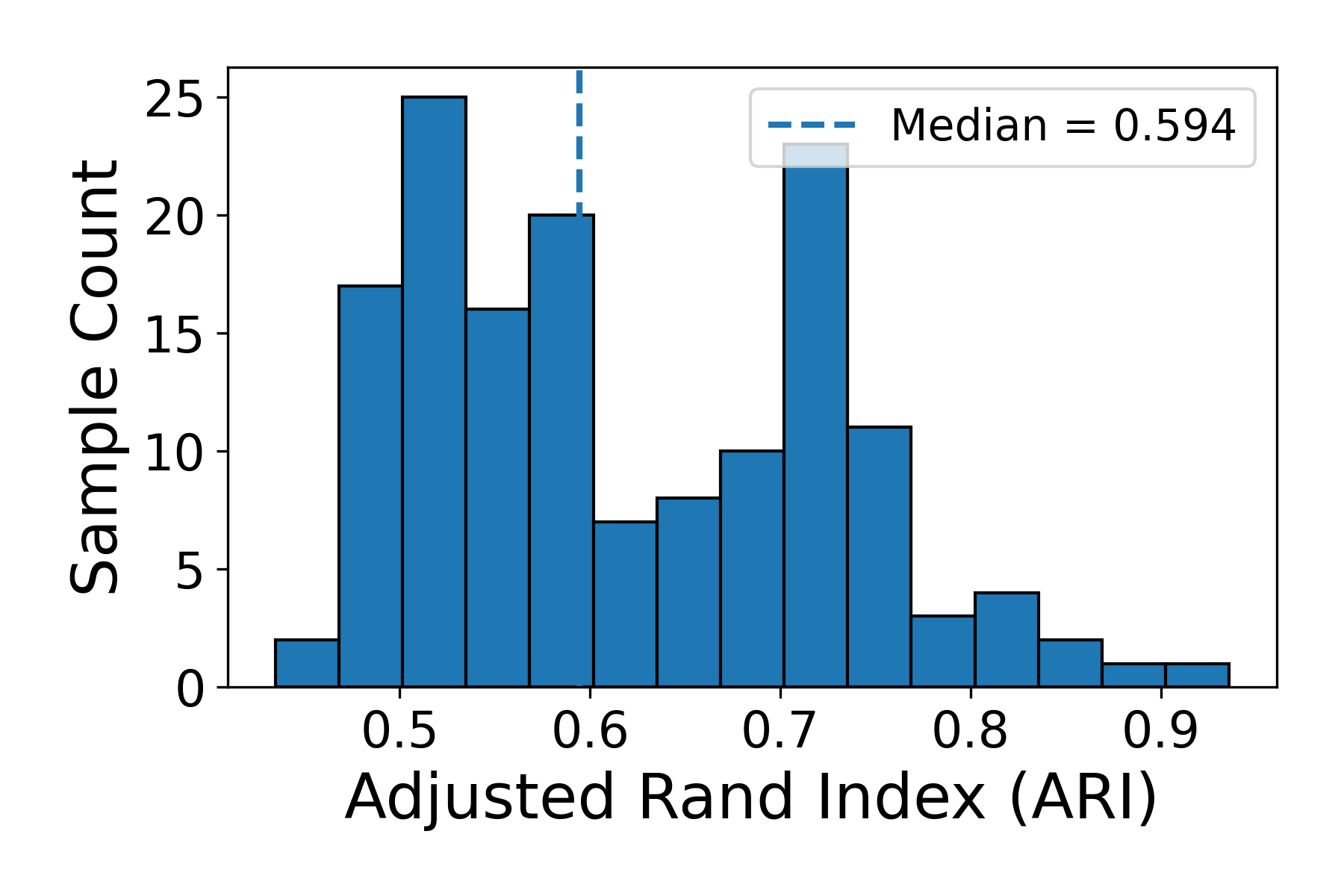}
\caption{Bipolar Disorder}
\label{fig:ari_stability_bipolar}
\end{subfigure}\hfill
\begin{subfigure}{0.50\linewidth}
\centering
\includegraphics[width=\linewidth]{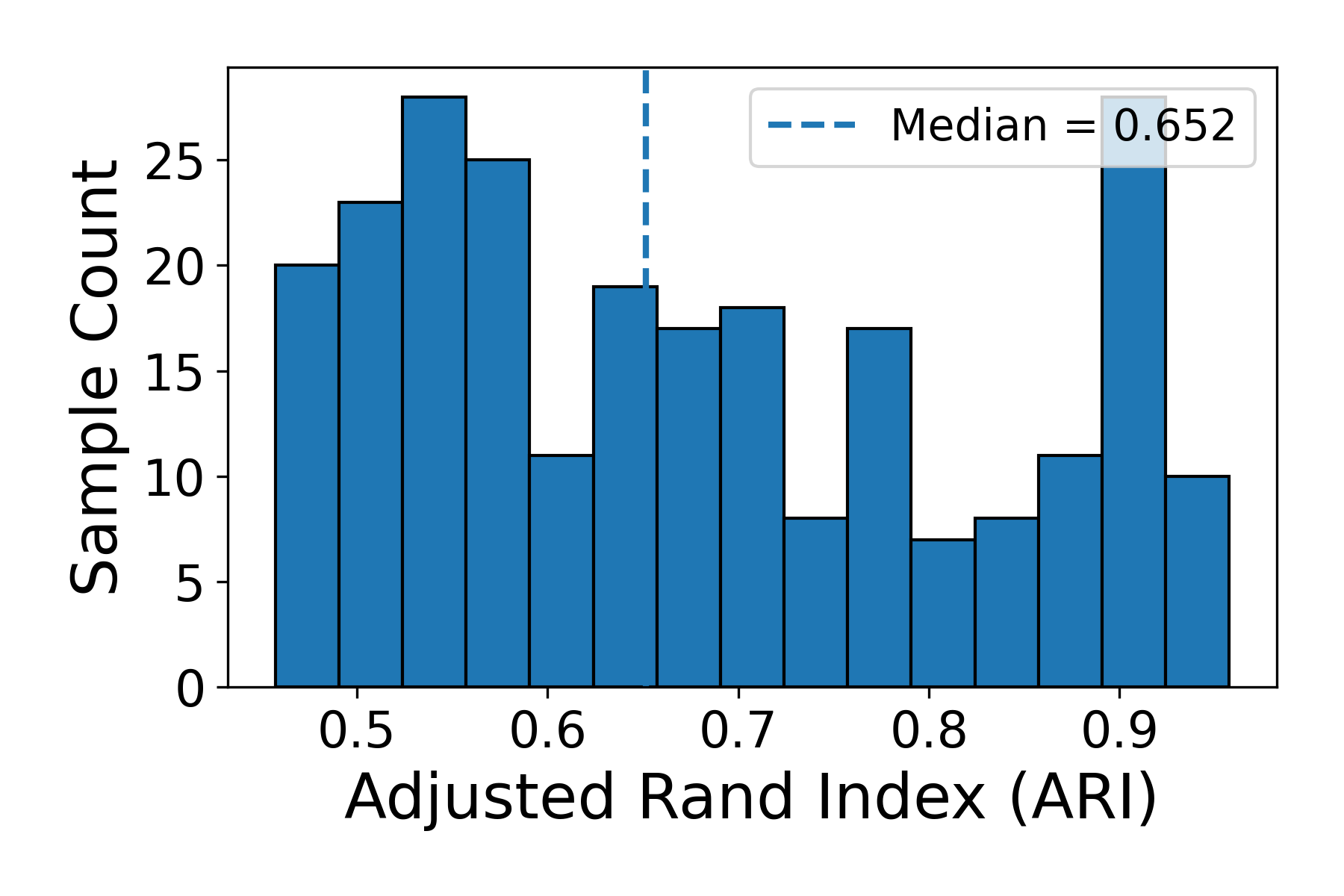}
\caption{Schizophrenia}
\label{fig:ari_stability_schizophrenia}
\end{subfigure}

\smallskip

\begin{subfigure}{0.50\linewidth}
\centering
\includegraphics[width=\linewidth]{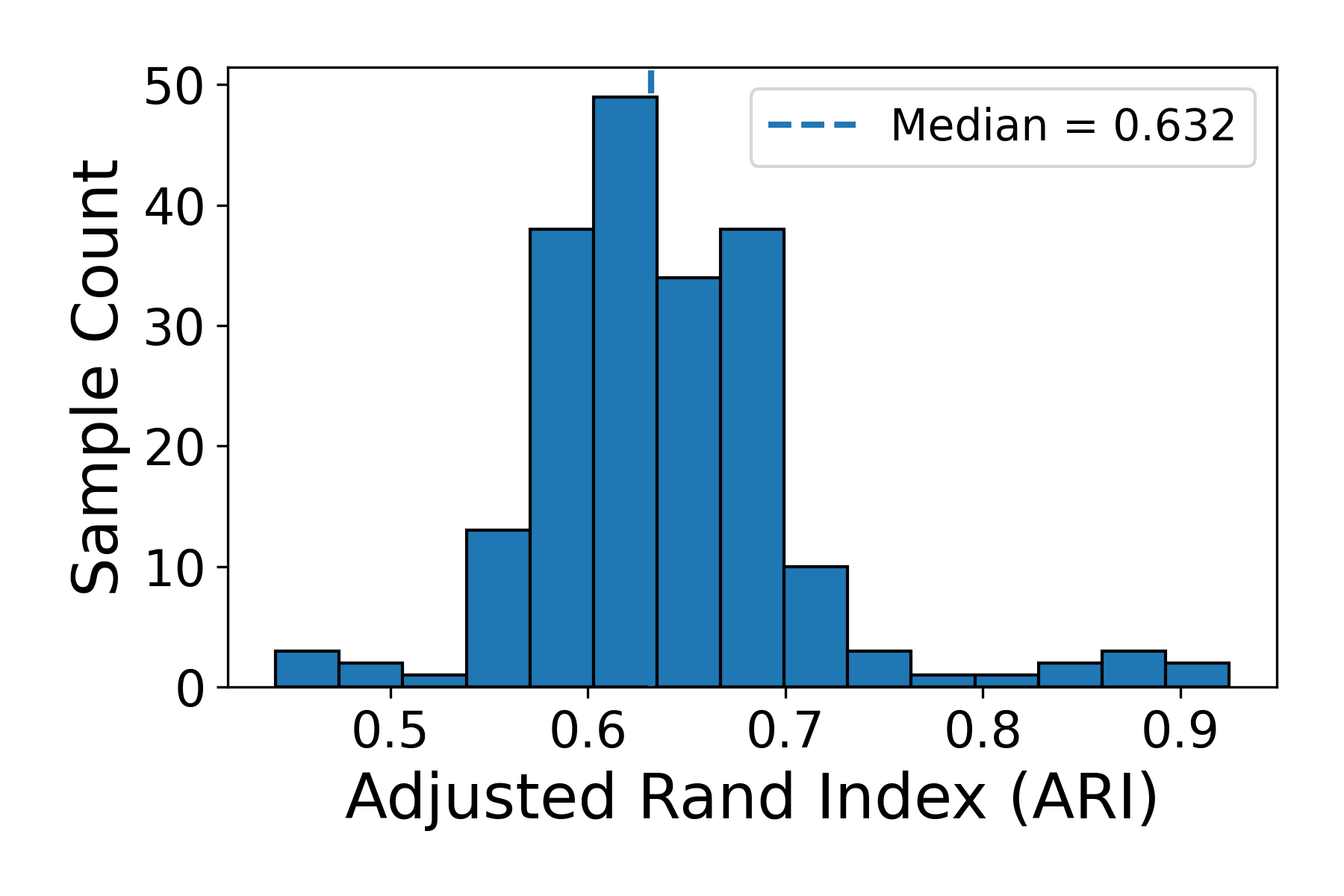}
\caption{Leukemia}
\label{fig:ari_stability_leukemia}
\end{subfigure}\hfill
\begin{subfigure}{0.50\linewidth}
\centering
\includegraphics[width=\linewidth]{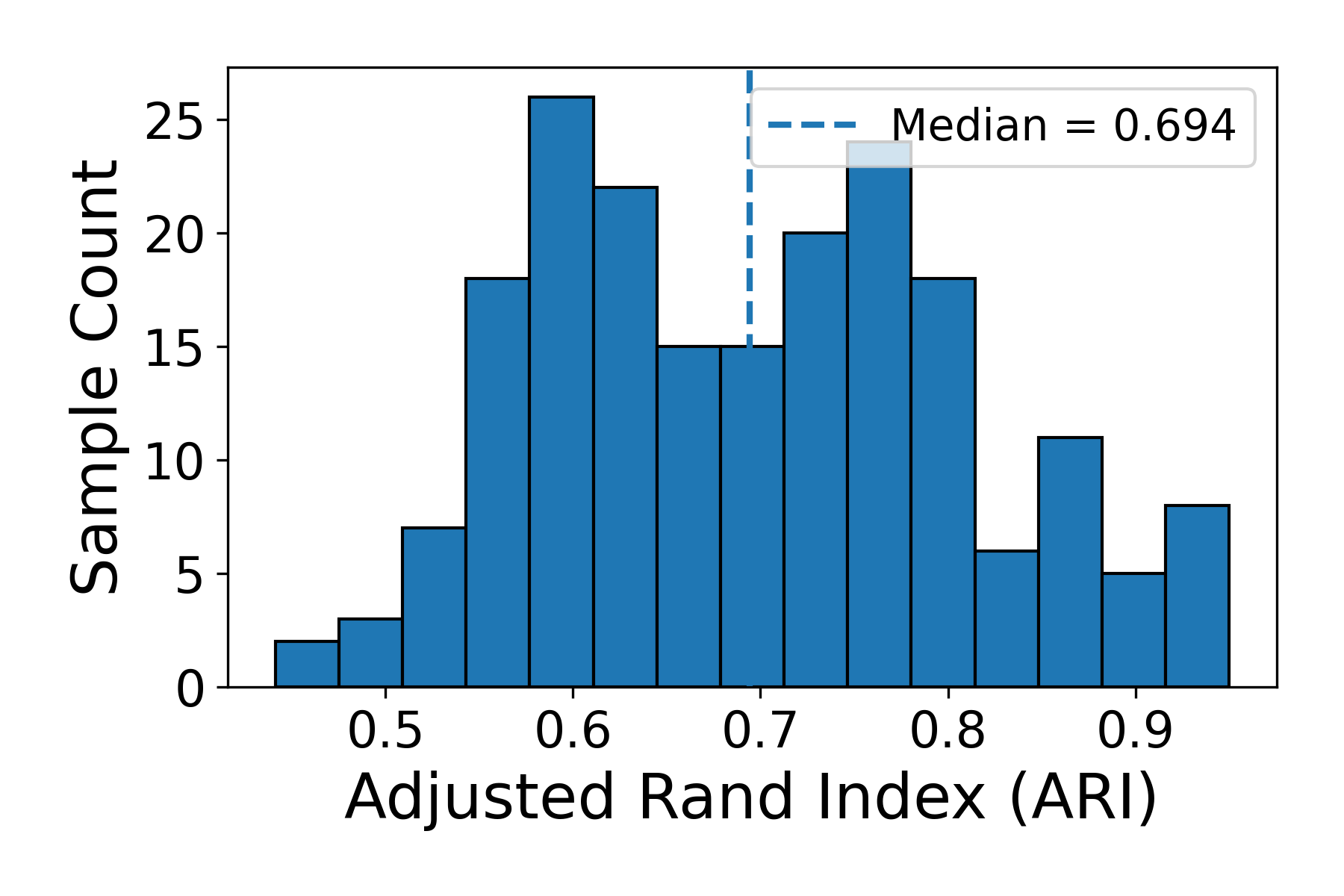}
\caption{Breast Cancer}
\label{fig:ari_stability_breastcancer}
\end{subfigure}

\medskip

\begin{subfigure}{0.50\linewidth}
\centering
\includegraphics[width=\linewidth]{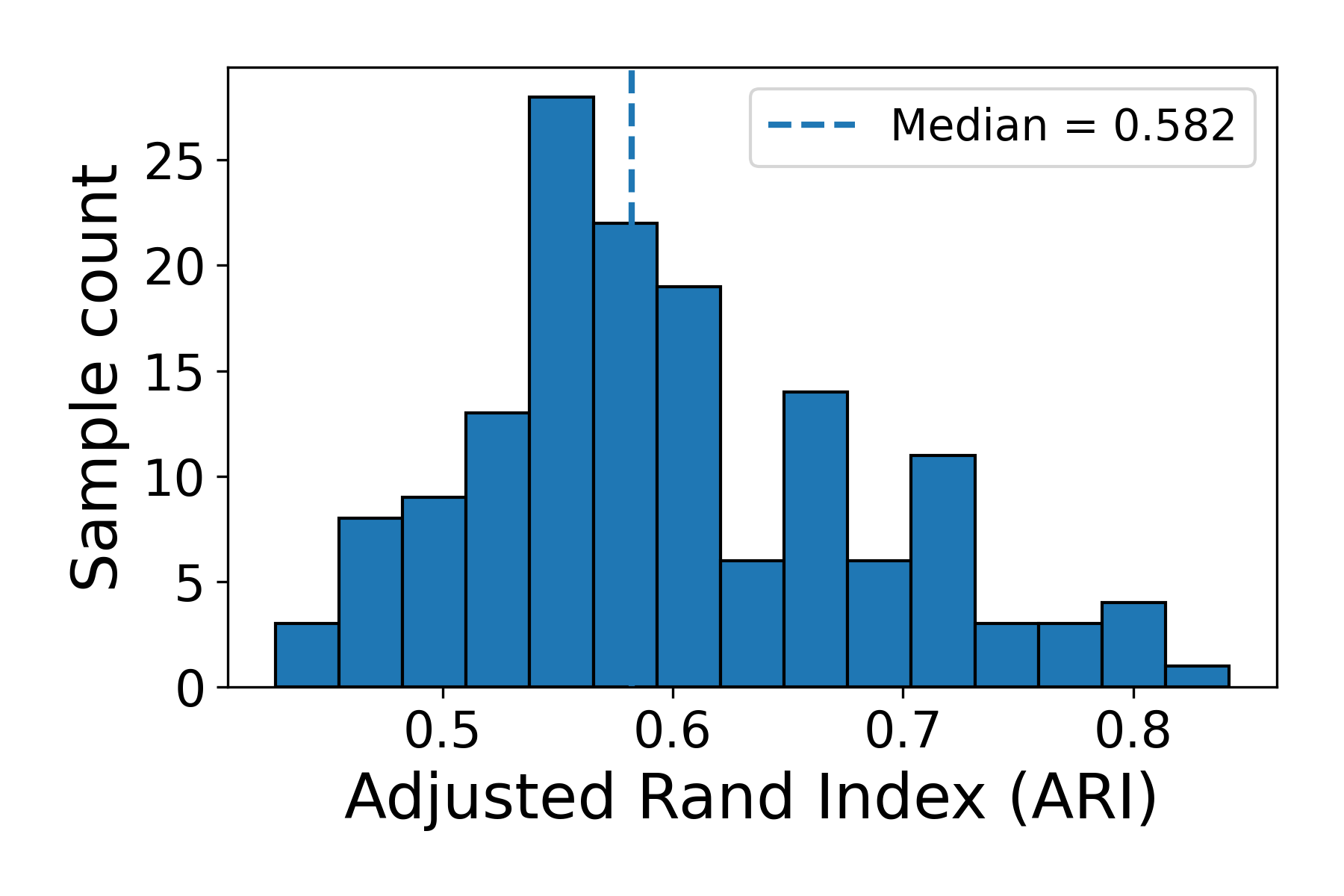}
\caption{MSigDB-only baseline}
\label{fig:ari_stability_baseline}
\end{subfigure}

\caption{Adjusted Rand Index stability distributions for disease-conditioned
multilayer hypergraphs and the MSigDB-only baseline. Each run samples \(80\%\)
of genes, reruns community detection on the induced graph, and compares the
resulting partition to the restricted full-data partition. The disease-conditioned distributions show stability comparable to, or slightly higher than, the baseline.
\newline
\textbf{Alt text}: Histograms show sample counts across adjusted Rand index values. Leukemia is concentrated around 0.6–0.7, while schizophrenia spans a broader range. Bipolar disorder and breast cancer have two prominent peaks; the baseline peaks near 0.55. Dashed lines mark medians, highest for breast cancer (0.694) and lowest for the baseline (0.582).
}

\label{fig:ari_stabilities}
\end{figure}

\endgroup

\subsection{Algorithm pseudocode}
\begin{algorithm}[H]
\caption{Sparse multilayer hypergraph diffusion clustering
}
\label{algorithm}
\begin{algorithmic}[1]

\State \textbf{Input}: Disease $d$, parameters outlined in Section 4

\State Form $\mathbb{S}$, the set of drugs associated with $d$ according to DDDB
\State Construct DGIdb hypergraph based on $\mathbb{S}$, weighting nodes using HumanNet
\State Construct universal MSigDB hypergraph, weighting nodes using HumanNet
\State Build \(A^{(D)},A^{(M)}\)
\State Build coupled operator \(P\)
\State Build aggregation matrices $A_r$ and $A_c$
\State Compute $\tilde{P}^t$ for \(t\in T\) (without forming $\tilde{P}$)
\State Build average squared diffusion distance matrix $\bar{D}^2$
\State  Build symmetrized \(k\)NN similarity graph $G$
\State Run weighted Leiden on $G$
\State Perform post hoc enrichment analysis.

\end{algorithmic}
\end{algorithm}
\paragraph{Run time: }We report wall-clock runtime measured on a 13th Gen Intel(R) Core(TM) i9-13900H CPU with 32GB RAM: Algorithm \ref{algorithm} takes roughly 536 minutes to run for a given disease with the parameter values given in Section 4. The corresponding coupled networks contain on the order of \(2\times 10^4\) unique genes and \(10^4\) hyperedges, leading to diffusion computations on tens of thousands of layer-specific supra-nodes. This runtime reflects an unoptimized CPU implementation; substantial speedups should be possible using GPU acceleration, low-rank diffusion approximations, approximate nearest-neighbor graph constructions, and smaller task-specific gene-set collections.

\bibliographystyle{plainnat}
\bibliography{references_final}
\end{document}